\pdfoutput=1
\documentclass[prd,aps,nofootinbib,twocolumn,superscriptaddress,preprintnumbers,balancelastpage,longbibliography]{revtex4-1}
\usepackage{graphicx}	
\usepackage{amsmath}	
\usepackage{dcolumn}
\usepackage{bm}
\usepackage{graphics}
\usepackage{afterpage}
\usepackage{float}
\usepackage{subfigure}
\usepackage{rotating}
\usepackage{multirow}
\usepackage{tabularx}
\usepackage{booktabs}
\usepackage{multirow}
\usepackage{fancyhdr}
\usepackage{hyperref}
\usepackage[utf8]{inputenc}
\usepackage{theorem}
\usepackage{moreverb}
\usepackage{euscript}
\usepackage{psfrag}
\usepackage{slashed}
\usepackage{mathtools}
\usepackage{makecell}
\usepackage[flushleft]{threeparttable}
\usepackage{adjustbox}
\usepackage{dcolumn}
\usepackage{bm}
\usepackage[mathlines]{lineno}
\usepackage{lettrine}
\usepackage[dvipsnames]{xcolor}
\usepackage{blindtext}
\usepackage{soul}

\input Zallman.fd

\LettrineTextFont{\itshape}
\hypersetup{
     colorlinks   = true,
     citecolor    = orange,
     urlcolor     = orange,
     linkcolor    = orange}
     
\usepackage{listings}
\usepackage{color,xcolor}

\newcommand{\appropto}{\mathrel{\vcenter{
  \offinterlineskip\halign{\hfil$##$\cr
    \propto\cr\noalign{\kern2pt}\sim\cr\noalign{\kern-2pt}}}}}

\begin{document}

\title{Dark Drag Around Sagittarius A$^\star$}
\author{Javier F. Acevedo}
\thanks{ \href{mailto:jfacev@uvic.ca}{jfacev@uvic.ca}; \href{https://orcid.org/0000-0003-3666-0951}{0000-0003-3666-0951}}
\affiliation{Particle Theory Group, SLAC National Accelerator Laboratory, Stanford, CA 94035, USA}
\affiliation{Department of Physics and Astronomy, University of Victoria, Victoria, BC V8P 5C2, Canada}
\affiliation{TRIUMF, 4004 Wesbrook Mall, Vancouver, BC V6T 2A3, Canada}

\author{Aidan J. Reilly}
\thanks{ \href{mailto:areilly8@stanford.edu}{areilly8@stanford.edu}; \href{https://orcid.org/0009-0005-4810-8920}{0009-0005-4810-8920}}
\affiliation{Particle Theory Group, SLAC National Accelerator Laboratory, Stanford, CA 94035, USA}

\author{Lillian Santos-Olmsted}
\thanks{\href{mailto:rleane@slac.stanford.edu}{solmsted@stanford.edu}; \href{https://orcid.org/0000-0002-8763-3702}{0000-0002-8763-3702}}
\affiliation{Particle Theory Group, SLAC National Accelerator Laboratory, Stanford, CA 94035, USA}
\affiliation{Kavli Institute for Particle Astrophysics and Cosmology, Stanford University, Stanford, CA 94305, USA}

\begin{abstract}
We analyze the effect of Dark Matter (DM) $-$ Standard Model (SM) non-gravitational interactions on the orbital dynamics of celestial bodies near the supermassive black hole Sagittarius A$^\star$, where the DM density is  generically expected to be high. We outline the conditions under which a DM–SM scattering channel gives rise to a drag force on objects in this region, and show that for sufficiently large cross-sections, this effect can lead to observable orbital decay on timescales as short as a single orbital period. We identify the types of objects most strongly affected by this dark drag and compute sensitivity to specific dark matter distributions and interaction strengths, assuming both elastic and inelastic scattering. For inelastic DM, we find sensitivity to mass splittings that reach near the MeV scale. We also demonstrate that a DM-induced drag force could potentially contribute to the observed depletion of red giant branch stars in the innermost region of the Milky Way. 
\end{abstract}

\maketitle

\section{Introduction}
The Galactic Center is one of the most extreme environments within the Milky Way, hosting the supermassive black hole Sagittarius A$^\star$ (Sgr.~A$^\star$) embedded in a dense population of stars, gas, and radiation. It is also a prime region to study scattering interactions between dark matter (DM) and the Standard Model (SM), considering the DM density there is generically expected to be orders of magnitude higher compared to anywhere else in the Milky Way. Motivated by this, several prior works have focused on DM capture and annihilation in celestial objects within this region \cite{Bramante:2014zca, Bramante:2015dfa, Bramante:2017ulk, 2019ApJ...879...50L, Lopes:2019kxo, Leane:2021ihh, Bose:2021yhz, Nguyen:2022zwb, Acevedo:2023xnu, John:2023knt, Bhattacharya:2023stq, Acevedo:2024ttq, John:2024thz, Acevedo:2024zkg, Linden:2024uph, Phoroutan-Mehr:2025hjz}. In contrast, less attention has been devoted to the dynamical effects that DM could have on the population of objects near Sgr.~A$^\star$, independent of capture and annihilation processes. The studies that do consider the orbital dynamics of Galactic Center objects typically only examine the gravitational effects of large dark matter densities in that region (see \emph{e.g.} \cite{Lacroix:2018zmg, Shen:2023kkm, Becerra-Vergara:2020xoj, Tomaselli:2025zdo}). While these have the benefit of constraining the dark matter profile close to Sgr.~A$^{\star}$ independent of any interaction, they are also limited by the dominating gravitational pull of Sgr.~A$^{\star}$.

In this work, we show how DM frequently scattering off the SM constituents of celestial objects in this region can give rise to a `dark drag' force on orbiting stars, which, over time, causes their orbital decay $-$an effect analogous to the atmospheric drag that orbiting satellites experience on Earth. The dissipative nature of such a force can lead to qualitatively new constraints on the dark matter profile and interaction strengths. We analyze the conditions under which this effect can be significant and its observational prospects. Previous studies have considered the effect of a cumulative force built up from dark matter scattering, usually in the context of earth-based experiments, using a net dark matter `wind' which, in the rest frame of the dark matter, takes the form of a drag force acting on the moving experiment \cite{Matsumoto:2025rcz, Gan:2025nlu, Fukuda:2021drn, Luo:2024ocg, Day:2023mkb}. While the experimental reach of these scenarios can be quite strong, they require precise instrumentation and rely on coherently enhanced scattering by low mass dark matter. On the astrophysical side, others have considered the effect of a dark matter induced drag force on planetary ephemeris in the solar system \cite{Fukuda:2018omk}. Such observations benefit from good astrometric precision of local objects, but are limited by the low density of dark matter at local position. By considering the effect of a DM-induced drag near the Galactic Center we can, in principle, probe smaller couplings. In particular, we show that objects in this region could experience sizable orbital changes relative to the dissipation-free case, after each pericenter passage, on timescales as short as a single orbit. We further show that red giant stars, owing to their large cross-sectional area, are especially susceptible to dark drag, making this mechanism a natural candidate for explaining the reported lack of giants within the innermost tenth of a parsec \cite{Schodel:2007er, 2009ApJ...703.1323D, 2009A&A...499..483B, 2010ApJ...708..834B, 2018A&A...609A..26G, 2019ApJ...872L..15H}. 

This paper is divided as follows: in Section~\ref{sec:dm_GC_dist} we review the expected dark matter distribution in the central region of the Galaxy. In Section~\ref{sec:drag_main} we review the theory behind the dark drag force, and analyze its various regimes depending on the interaction strength and DM mass. In Section~\ref{sec:drag_targets} we discuss the short timescale observational prospects of this effect by considering the orbital evolution of a sample of stars near Sgr.~A$^\star$ (known as the S-cluster) \cite{2017ApJ...837...30G}, as well as the Galactic Center gas cloud G2 \cite{gillessen2012gas}. Intriguingly, for the latter object there are indications of a drag force acting on it. In Section~\ref{sec:cs_general}, we examine the constraints on a DM-SM portal that can be inferred from G2's orbital evolution. In Section~\ref{sec:miss_giants}, we discuss how a dark drag force could potentially be connected to the well-established dearth of giants in the innermost region of the Galactic Center. We conclude in Section~\ref{sec:conclusion} by summarizing our findings and stating future directions. Unless otherwise specified, we take natural units $c = \hbar = k_b = 1$ and $G = M^{-2}_{\rm Planck}$ throughout this work.

\section{Dark Matter at the Galactic Center}
\label{sec:dm_GC_dist}
As our goal is to understand the dynamical effects that the high dark matter density at the Galactic Center has on stars in the presence of non-gravitational DM-SM portals, we will consider a variety of distributions in order to span the existing astrophysical uncertainties. We will primarily focus on power-law profiles, which may or not include a DM spike. DM spikes are a well-motivated possibility based on an adiabatic growth history of Sgr.~A$^\star$ \cite{Gondolo:1999ef} (although see \cite{Ullio:2001fb, Merritt:2002vj, Bertone:2024wbn}). In this work, we remain agnostic about the existence of a spike, and present results for both the more conservative scenario of a cuspy Navarro-Frenk-White (NFW) profile \cite{Navarro:1995iw, Navarro:1996gj}, 
\begin{equation}
\label{eq: nfw density}
    \rho^{\rm (NFW)}_\chi(R) = \frac{\rho_\chi^0}{\left(\frac{R}{R_s}\right)^\gamma\left(1+\left(\frac{R}{R_s}\right)\right)^{3-\gamma}}~,
\end{equation}
valid down to $ R \geq 2R_{\rm Sg A^\star}$, and also consider the more aggressive case of a spike,
\begin{equation}
\label{eq: spike density}
    \rho_{\chi}^{\rm (sp)}(R) = \begin{cases}
        0 \qquad\qquad\qquad &R< 2R_{\rm SgA^{*}}\\
        \rho_\chi^{(\rm NFW)}(R_{\rm sp}) \left(\frac{R}{R_{\rm sp}}\right)^{-\gamma_{\rm sp}} \quad &2R_{\rm SgA^{*}}\leq R<R_{\rm sp}\\
        \rho_\chi^{\rm (NFW)}(R)  & R_{\rm sp} \leq R
    \end{cases}~.
\end{equation}
Above, $R_{\rm SgA^{*}}$ is Sgr.~A$^{\star}$'s Schwarzschild radius, $\gamma$ is the NFW slope, taken in the range $1.0 - 1.5$ motivated by adiabatic contraction analyses~\cite{2011arXiv1108.5736G,DiCintio:2014xia}, and $R_s$ is the scale radius which we fix to $R_s = 12 \ \rm kpc$. We refer to the NFW profile with $\gamma = 1$ as just NFW, and $\gamma = 1.5$ as generalized NFW (gNFW) profile. The normalization $\rho_\chi^0$ is fixed so that $\rho_\chi \simeq 0.42 \ \rm GeV \ cm^{-3}$ at the heliocentric position. On the other hand, $R_{\rm sp}$ is the spike radius and $\gamma_{\rm sp}$ is the spike index, for which we consider two possibilities. The first spike index is calculated according to the Gondolo-Silk model (GS spike) \cite{Gondolo:1999ef},
\begin{equation}
    \gamma_{\rm sp}^{(\rm GS)} = \frac{9-2\gamma}{4-\gamma}~,
\end{equation}
where $\gamma$ is the index of an initial NFW profile. The other spike index we consider incorporates the possibility of significant softening due to gravitational heating by the nuclear star cluster \cite{Gnedin:2003rj, Bertone:2005hw, Merritt:2006mt, Shapiro:2022prq}, leading to an equilibirium density index of 
\begin{equation}
    \gamma_{\rm sp}^{\rm heated} = 1.5~.
\end{equation}
Numerical studies indicate that the spike begins to grow around $R_{\rm sp} \simeq 0.34$ pc \cite{Balaji:2023hmy}, while earlier studies suggest larger spike radii of $R_{\rm sp}\simeq 18$ pc \cite{Gondolo:1999ef}, and observations constrain $R_{\rm sp}$ at a level somewhere between the two \cite{Shen:2023kkm}. Therefore, we choose a fiducial value of $R_{\rm sp} = 1$ pc for all spike profiles. Finally, while it is entirely possible for a spike to grow on top of any DM profile, we only consider spike scenarios with an initial NFW profle of $\gamma = 1$ (for a more in depth review of dark matter spike possibilities, see \cite{Balaji:2023hmy}).

\begin{figure}[t]
    \centering
\includegraphics[width=\linewidth]{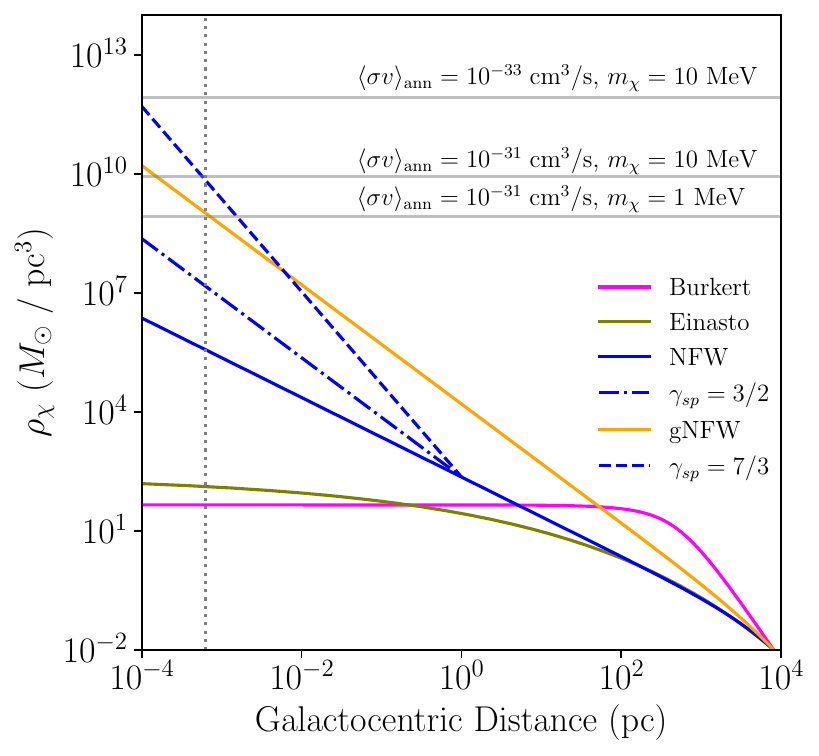}
        \caption{Various dark matter density profiles considered in this work. Lines labeled as NFW and gNFW correspond to Eq.~\eqref{eq: nfw density} with slopes of $\gamma = 1$ and $\gamma = 1.5$ respectively. Lines labeled by their spike index $\gamma_{\rm sp}$ correspond to DM spikes which grow on top of an NFW profile with a spike radius of $R_{\rm sp}=1$ pc (see text). Burkert and Einasto refer to the cored profiles defined by Eqs.~\eqref{eq: Burkert} and \eqref{eq: Einasto}. The vertical dashed line corresponds to $r=131.5$ AU, the distance of gas cloud G2's last pericenter passage for reference~\cite{gillessen2012gas}. The gray horizontal lines show $\rho^{\rm (max)}_\chi$ for symmetric DM with various masses and annihilation cross-sections.}
    \label{fig: density profiles}
\end{figure}

We additionally include in our analysis the cored Burkert \cite{Pato:2015dua} and Einasto profiles \cite{Graham:2005xx, Pieri:2009je}, in order to further span the theoretical possibilities for the DM distribution in this region. These are respectively given by
\begin{equation}
\label{eq: Burkert}
    \rho^{\rm (Bur)}_\chi(R) = \frac{\rho_\chi^0}{\left(1 + \left(\frac{R}{R_{\rm sb}}\right)\right)\left(1 + \left(\frac{R}{R_{\rm sb}}\right)^2\right)}~,
\end{equation}
and
\begin{equation}
\label{eq: Einasto}
  \rho^{\rm (Ein)}_\chi(R) = \rho_\chi^0 \, \exp{\left[-\frac{2}{\beta}\left(\left(\frac{R}{R_s}\right)^{\beta}-1\right)\right]}~. 
\end{equation}
Above, we fix the Burkert core radius $R_{\rm sb} = 0.5 \ \rm kpc$ \cite{Cohen:2013ama}, whereas for Einasto we fix $R_s = 20 \ \rm kpc$ and $\beta = 0.17$ \cite{Pieri:2009je}. As before, their normalization is fixed so that $\rho_\chi \simeq 0.42 \ \rm GeV \ cm^{-3}$ at the heliocentric position. For such profiles, however, we do not find significant cycle-to-cycle orbital changes for any known objects. 


The above distributions are technically valid for asymmetric or non-annihilating DM, in which case annihilation is suppressed and the DM density can grow unchecked toward $R \rightarrow 0^{+}$. For symmetric DM, however, annihilation leads to a maximum central density determined by its cross-section $\langle \sigma v \rangle_{\rm ann}$ \cite{laTorreLuquePedro:2024est}
\begin{equation}
    \rho_\chi^{(\max)} \simeq \frac{m_{\chi}}{\langle \sigma v \rangle_{\rm ann} t_{\rm{BH}}}, 
\end{equation}
where $m_{\chi}$ is the DM mass and $t_{\rm{BH}} \simeq 10^{10}$ yr is the age of Sgr. A$^{\star}$. While strictly speaking we will consider just the asymmetric case, only strong annihilations will affect the DM distribution at the distance of closest approach for the objects we examine. 
Figure~\ref{fig: density profiles} illustrates these various DM profiles for the inner 10 kpc for various parameters as specified. To address the annihilating DM case, we also include lines in Fig.~\ref{fig: density profiles} corresponding to the maximum DM density for different values of DM mass and annihilation cross section.

\section{Dark Matter Induced Drag Force}
\label{sec:drag_main}
We now begin by calculating the drag force that ambient dark matter will impart upon on a celestial body orbiting near the Galactic Center. We will work in the limit of no or weak DM self-interactions. This choice allows for a simplified calculation of the drag coefficient from first principles, and only requires the mean free path of the dark matter ``gas" to be larger than the size of the object in question, quantifying the so called ``free molecular flow limit" (see App.~\ref{app: Drag Force} for details). Moreover, this requirement is met in most scenarios, even for moderately self-interacting dark matter. However, we note that as the dark matter mass is reduced, the number density goes up accordingly when holding the mass density constant. Because the self-scattering probability is a function of number density, this has consequences for the assumption that self-scattering can be ignored, and implies certain limits on the self-interaction cross section. While we maintain the assumption of working in the free molecular flow limit for all masses, we also emphasize that we do not expect the constraints set to disappear once self-interactions are taken into account. Typical drag forces in the hydrodynamic limit only differ from those in the free molecular flow limit by an ${\cal O}(1)$ factor known as the drag coefficient. We do not include this scenario in our analysis because this ${\cal O}(1)$ factor can typically only be determined with experiment or simulation. 
We therefore leave a determination of the drag coefficient in other regimes to future work. 

\subsection{Total Reflection}
\label{subsec: drag total reflection}
The first scenario we consider involves a calculation of the drag force when every incident particle is isotropically scattered off the body in question. This is the limit that typically applies to SM-SM scattering, but will be cross section-dependent with DM-SM scattering. When the size of the impinging object is much larger than individual DM particles, a condition met by many orders of magnitude for all DM models that we consider, the drag force takes the form \cite{Hoang:2017ijx, Baines:1965, Draine1979} 
\begin{equation}
\label{eq: Fdrag full}
    \vec{F}_{\rm drag}^{\rm geo} = 2\pi R_\star^2n_\chi \left[T_\chi \, \frac{8v^2}{3\sqrt{\pi}v_\chi^2}\left(\left(\frac{v_\chi}{v}\right)^2+\frac{9\pi}{64}\right)^{1/2}\right] \, \hat{v}~ \,,
\end{equation}
for a Maxwellian DM velocity distribution of temperature $T_\chi$ and thermal velocity $v_\chi = \sqrt{2T_\chi/m_\chi}$. In Eq. \eqref{eq: Fdrag full}, $R_\star$ is the radius of the celestial body, $v$ is its velocity (with $\hat v$ its associated unit vector), and $n_\chi$ is the DM number density at the object's position. In the limit that $v \gg v_\chi$, which is an assumption we make moving forward, Eq.~\eqref{eq: Fdrag full} takes the simple form: 
\begin{equation}
\label{eq: Fdrag simple}
\vec{F}_{\rm drag}^{\rm geo} \simeq  -\rho_\chi \pi R_\star^2 v^2 \, \hat{v}~,
\end{equation}
with $\rho_\chi = m_\chi n_\chi$ the DM mass density. This choice not only simplifies calculations, but is also a conservative method to incorporate our relative ignorance about the DM velocity distribution at the galactic center. Indeed, it is trivial to verify that Eq.~\eqref{eq: Fdrag full} $\geq$ Eq.~\eqref{eq: Fdrag simple} for all values of $v$ and $T_\chi$ 
. This approximation is further justified because the pericenter velocity for most objects closely orbiting Sgr.~A$^\star$ is larger by a factor of a few than the inferred DM velocity dispersion at the same distance \cite{2013PASJ...65..118S}.  We have verified that incorporating the effects of the non-zero DM velocity dispersion has a mild $\mathcal{O}(1)$ impact on the orbital decay times for the objects we consider here. 


\subsection{Partial Reflection}
\label{subsec: drag partial reflection}
When the probability that an impinging DM particle will scatter with the celestial body cannot be approximated to 1, we must re-write Eq.~\eqref{eq: Fdrag simple}. It is still true that the full celestial body will always be more massive than individual dark matter particles, but for low scattering cross sections we cannot assume that the DM particle scatters off of the entire body. Instead, a DM particle is likely to only scatter one time (for the cross sections we will work with), and therefore the transferred momentum is a function of the ratio $m_\chi/m_{T}$, with $m_{T}$ the mass of the constituent SM target. This is the regime in which the mean free path of DM through the object in question is larger than the object's overall size. Note that we will continue to work in the limit that the velocity of the surrounding dark matter is negligible even in the long mean free path regime (see App. \ref{app: DM velocity} for a discussion of why this assumption remains reasonable and conservative).

In this regime, we can write the general drag force as 
\begin{equation}
\label{eq: Fdrag avg}
    \vec{F}_{\rm drag} = \langle n_\chi \, v \, A_{\rm eff} \, {\vec q}\rangle,
\end{equation} where $A_{\rm eff}$ is the effective area for scattering, now a function of the DM-SM interaction strength as well as the size of the celestial body, $n_{\chi}$ is the ambient number density of dark matter, $\vec{q}$ is the transferred momentum in a single scattering event, and $v$ is still the velocity magnitude of the object. As $n_\chi$ and $v$ do not change with each collision, these quantitites can be factored out of the average in Eq.~\eqref{eq: Fdrag avg}. Furthermore, when scattering is isotropic, $A_{\rm eff}$ is independent of momentum transfer and can be factored out as well. The average momentum exchange in a non-relativistic, isotropic, elastic collision of an incident dark matter particle with a SM target at rest is simply 
\begin{equation}
    \label{eq: q avg} 
    \langle \vec{q} \rangle = \mu_{\chi T} \;\vec{v}\ ,
\end{equation}
where $\mu_{\chi T}$ is the reduced mass the two particles.
Different limits of the dark matter mass thus lead to the drag force
\begin{equation}
    \label{eq: Fdrag eff}
    \vec{F}_{\rm drag} = \begin{cases} -\left(\frac{m_{T}}{m_\chi}\right)\rho_\chi A_{\rm eff} v^2 \, \hat{v}~, \quad m_\chi \gg m_{T} \\\\
    -\rho_\chi A_{\rm eff} v^2 \, \hat{v}~, \qquad \qquad  m_\chi 
    \ll m_{T}
    \end{cases},
\end{equation}
where we have used the fact that $n_\chi = \rho_\chi /m_\chi$. As one can see, the force drops off linearly with mass for $m_\chi > m_{T}$, but stays constant for $m_\chi < m_{T}$. Note that we have not included the transition regime where $m_\chi \simeq m_T$ in Eq. \eqref{eq: Fdrag eff}. In this case $\mu_{\chi T}\approx m_\chi/2$, and this factor of 1/2 appears in the magnitude of the drag force. What remains, then, is to calculate $A_{\rm eff}$. For most DM masses, the effective area will simply be 
\begin{equation}
\label{eq: A_eff}
    A_{\rm eff} \simeq N_{\rm SM}\;\sigma_{\chi T}~,
\end{equation}
where $N_{\rm SM}$ is the total number of SM scatterers in the object, and $\sigma_{\chi T}$ is the DM-SM interaction cross-section.  The exact form we use for $\sigma_{\chi T}$ in different mass regimes is described in Section \ref{sec:cs_general}, but we will always assume the mediating interaction between DM and the SM to be short-ranged. Eq.~\eqref{eq: A_eff} holds up until the point where $A_{\rm eff} = \pi R_\star^2$, at which point the geometric limit is saturated and one should use Eq. \eqref{eq: Fdrag simple}. 

When the DM is ultralight, coherent effects can change the scaling of $A_{\rm eff}$ with $N_{\rm SM}$, as discussed in App.~\ref{sec: ultralight dm}. There also exists a regime for heavy dark matter in which $A_{\rm eff} = \pi R_\star^2$, but because $\langle q\rangle < m_\chi v$, total reflection is not reached. In this case, one must calculate the probability for multiple scatters and the associated momentum loss to correctly extract the drag force. However, such a  regime is beyond the scope of our analysis, as we are primarily interested in light dark matter with cross sections low enough that at most one scatter occurs. 

Finally, while Eq. \eqref{eq: Fdrag eff} applies to isotropic scattering, one has to be careful when generalizing the formula to non-isotropic scattering. In this case, not only will the cross section be a function of energy, but so too will the average momentum transfer. Therefore, one must start with Eq. \eqref{eq: Fdrag avg} and take the appropriate average of the entire function. We discuss an example of such a scenario in more detail when considering inelastic DM in Sec. \ref{sec:heavy_inel_DM}.

\section{Dark Drag Targets}
\label{sec:drag_targets}
We now review the existing observations of some known Galactic Center objects, and discuss potential sensitivity to a DM-induced drag force.

\subsection{The S-cluster Stars}
Several stars in close proximity to Sgr.~A$^{\star}$ have well-measured orbits and large pericenter velocities \cite{2017ApJ...847..120H, 2017ApJ...837...30G}, making them potential probes of the dark drag effect analyzed here. For completeness, we examine a few representative examples, and estimate their sensitivity by comparing the uncertainty in their orbital periods with the energy loss due to the drag force. However, despite the high precision astrometric observations of these objects, we find that their sensitivity is not high enough to produce meaningful limits within a single orbital cycle. As such, S-cluster stars will not be used in subsequent analyses and this section can be skipped without loss of continuity.  \\

\textbf{S2.~$-$} S2 is one of the brightest and most extensively studied stars among Sgr.~A$^{\star}$'s cluster, with its orbital elements measured to astounding precision and providing crucial evidence supporting the massive black hole nature of Sgr.~A$^{\star}$ \cite{2017ApJ...837...30G, Do:2019txf}. It is a B0-2V main-sequence star with approximate mass $M_{\rm S2} \simeq 14 M_\odot $ and radius $R_{\rm S2} = 6.4 R_\odot$ \cite{2017ApJ...847..120H}. As a simple order-of-magnitude estimate, we simulated S2's evolution following \cite{Do:2019txf}. Given its reported eccentricity $e \simeq 0.885$ and period $P \simeq 16.04 \ \rm yr$, we find that only dark matter densities consistent with an extreme spike profile (or, alternatively, a highly-sloped gNFW profile with $\gamma = 1.8 - 1.9$) could cause an observable variation of its orbital period at the level of the existing uncertainty $\sim \pm 0.02 \ \rm yr$ in the next pericenter passage. Such high densities would also be in tension with limits on the extended mass around Sgr.~A$^\star$ from the GRAVITY collaboration \cite{2020A&A...636L...5G}. A full orbital refit of S2 including the dark drag, combined with even greater astrometric precision, could offer increased sensitivity for more moderate DM profiles in the future. 

\textbf{S38.~$-$} S38 is a G7-8III star on an orbital period of $19.2 \ \rm yr$ \cite{2017ApJ...837...30G}. Based on its most recent spectral classification, it has a mass and radius $M_{\rm S38} \ \simeq 6 \, M_\odot$ and $R_{\rm S38} = 4 \, R_{\odot}$ \cite{2019ApJ...872L..15H}, which implies that this star would experience a drag force at a similar level as S2 for the same cross section. However, unlike S2, its orbital parameters have been determined to a lesser degree of precision, and the longer period means fewer pericenter passages within a given observational timespan. Therefore, we expect its sensitivity prospects to be less promising than S2.

\textbf{S4714.~$-$} S4714 is a faint star of shorter period than S2 at $P \simeq 12 \ \rm yr$, and an extreme eccentricity at $e \simeq 0.985$ \cite{2020ApJ...899...50P}. Its considerably higher pericenter velocity compared to the above cases implies a substantially larger energy dissipation with each passage, even assuming masses and radii comparable to solar values or those of S4716 (see below). However, the current uncertainties in its orbital parameters associated with the difficulty in observing this star, in particular the period uncertainty of $\sim \pm 0.3 \ \rm yr$, limit the sensitivity achievable via the same simple estimate applied to S2 above. 

\textbf{S4716.~$-$} S4716 is currently the fastest known star on a $P \simeq 4 \ \rm yr$ orbital period \cite{2022ApJ...933...49P}. Despite its smaller inferred radius $R_{\rm S4716} \simeq 2.5 R_\odot$, the higher ambient dark matter density and orbital velocity for this star imply a potentially stronger orbital decay compared to S2. However, its orbital parameters are currently measured with much larger uncertainties, and higher-order relativistic effects must be included, complicating the analysis. Assuming further improvements in the astrometry of this object, future pericenter passages could provide enhanced sensitivity to a dark drag force, but we leave a full orbital fit and sensitivity forecasts for this star for future investigation.

\subsection{Galactic Center Source G2}
\label{subsec: G2 orbit}
Having established the sensitivity prospects for a sample of S-stars, we turn to the low-density object G2, which appears to be an ionized gas cloud on a highly eccentric orbit around Sgr.~A$^{\star}$. G2, first detected in \cite{gillessen2012gas}, passed pericenter in 2014 at a distance $r_p \approx 130 \text{ AU}$ (where are have adopted the orbital parameters of \cite{Gillessen:2019} for consistency). Its large size and close pericenter distance makes it particularly sensitive to drag forces, and indeed there is recent positive confirmation of a dissipative force acting on it. 

While the origin and exact composition of G2 is still a topic of active investigation (see e.g. \cite{plewa2017post} and the references therein), the extended nature of the emissions, line width, and apparent tidal stretching on its approach towards Sgr. A$^{\star}$ lead to the expectation that it consists either of a gaseous dust envelope containing a compact stellar core or of an independent gas/dust cloud \cite{Morsony:2015uqa, plewa2017post, Gillessen:2019}(see also recent observational evidence that G2 is part of a larger gas streamer \cite{2025arXiv251000897G}). Even if such a cloud does contain a stellar core, the gravitational dynamics of the cloud this close to Sgr.~A$^{\star}$ are dominated by the black hole potential, and it can be ignored in orbit fitting (for all reasonable mass estimates of the core). On the other hand, some argue for scenarios with no extended gas cloud, e.g. that G2 could include a stellar core with only a compact dust envelope \cite{Zajacek2014,Zajacek:2024jqp, peissker2021apparent}. The effect of a DM induced drag force in this case would qualitatively change, and we leave an exploration of orbital dynamics in such scenarios to future study, but work under the gas cloud assumption here. We furthermore make the simplifying assumption that the gas cloud can be modeled as a uniform sphere throughout its orbital evolution. While in reality the cloud tidally evolves, there exist, for example, magnetically arrested cloud scenarios that can stabilize the geometry to some extent \cite{Shcherbakov_2014,McCourt2015}. In Appendix \ref{app: G2 shape} we provide a discussion concerning the shape of G2 and how much the true shape can deviate from the spherical model before it affects our results. 

The total mass of the gaseous component can be estimated as
\begin{equation}
    M_{\rm G2} \simeq 3 \sqrt{f_V}\left(\frac{R_{\rm G2}}{1.82 \times 10^{15} \text{ cm}}\right)^{3/2}M_{\rm Earth}~,
\end{equation} 
where $M_{\rm Earth} \simeq 3 \times 10^{-6} \ M_{\odot}$ is the Earth's mass for reference, $R_{\rm G2}$ is its rough spherical radius before pericenter passing, and $f_V$ is the volume filling factor \cite{gillessen2012gas, pfuhl2015galactic}. We take $f_V \simeq 1$, and note that in the partial reflection regime, the drag force scales proportionally to the number of targets, and therefore the total mass. Because acceleration scales inversely with mass, the overall drag effect becomes independent of $f_V$. 

Such an extended object passing so close to the galactic center is an ideal probe of drag forces induced by DM interactions. In fact, G2's orbit has been used to constrain the unknown SM density at its pericenter using exactly the same effect. Ref.~\cite{Gillessen:2019} found that G2's orbit is best fit by including a small drag force experienced during its pericenter passage, compatible with the limits set by previous studies \cite{pfuhl2015galactic,plewa2017post}.  We remark that neither the Keplerian nor the drag fit of the orbit result in a small enough $\chi ^2$ to be a true fit, but the drag orbit provides a significant improvement. The orbit fitting procedure involves a test particle model, which breaks the cloud into a set of N clumps, each following their own orbit around Sgr.~A$^{\star}$, neglecting self-interactions of the cloud. To account for potential drag, a velocity and distance dependent acceleration is considered of the form
\begin{equation}
\label{eq: Fdrag SM}
    \vec{a}_D = -c_D\; r^{-\alpha}\; v^2 \, \hat{v}~.
\end{equation}
The parameter $c_D$ is then fit to match the gas cloud emissions over its observed orbital evolution, for a given $\alpha$, which parameterizes the assumed distribution of SM matter in Sgr.~A$^\star$'s accretion disk, exactly analagous to the parameter $\gamma$ in the DM density profiles of Sec. \ref{sec:dm_GC_dist} (note that for $\alpha = 1$, $c_D$ is dimensionless). This method has the benefit of naturally encompassing projection effects which can impact the appearance of G2, and was shown to reproduce observations relatively well \cite{plewa2017post}. Ref.~\cite{Gillessen:2019} also used an even simpler centroid model where the cloud is treated as a single point source and achieved  qualitatively similar results, so we follow the simplified assumptions of the centroid model moving forward (this is is also the more conservative choice as the bounds are weaker). Furthermore, the authors of \cite{Gillessen:2019} find that varying the slope of the density profile between $\alpha \in (0.5, 1.5)$ introduces an additional factor of $\sim$ 4 uncertainty on the observed drag force at pericenter. For the density profiles that we consider, this factor of 4 is extremely minor relative to the multiple orders of magnitude  difference in pericenter densities that the different slopes correspond to.
Consequently, the results are not very sensitive to the specific choice of $\alpha$, and we will therefore bound only the drag induced acceleration at pericenter $\vec{a}_D(r_p)$, treating the choice of $\alpha$ as uncertainty in this constraint. This approximation is conservative for all profiles except for the Gondolo-Silk spike which has a power law scaling in density outside of the range considered in \cite{Gillessen:2019}. We therefore caution that there is additional uncertainty in the quoted results for that profile.

Following the procedure of \cite{Gillessen:2019} where Eq.~\eqref{eq: Fdrag SM} is mapped onto a ram-pressure drag formula, we can use the maximum derived SM density to constrain $\vec{a}_D(r_p)$. We will thus approximate an upper limit on the magnitude of $\vec{a}_D(r_p)$ for all density profiles as
\begin{equation}
    \label{eq: a_D(r_p)} 
    |\vec{a}_D(r_p)| \leq \frac{m_n}{M_{\rm G2}}n_0 \frac{\pi}{2}R_{\rm G2}^2 v_p ^2~,
\end{equation}
to draw constraints in Section \ref{sec:cs_general}. Here we take $R_{\rm G2} = 1.82 \times 10^{15}$ cm, $v_p \simeq 0.02$ is G2's velocity at pericenter, $m_n$ is the nucleon mass, and $n_0 = 8\times 10^{3} \text{ cm}^{-3}$ is the maximum SM number density found to be compatible with G2's orbital observations, normalized at 1000 Schwarzschild radii \cite{Gillessen:2019}. Depending on the value of $\alpha$, the density at $r_p \approx 1500$ Schwarzschild radii will be lower than $n_0$ by at most a factor of a few, which we conservatively neglect.
In principle, one should run the same analyses across all DM profiles of interest. However, as the results are not very sensitive to $\alpha$ and because observations of G2 already carry large uncertainties, we will use Eq. \eqref{eq: a_D(r_p)} to set appropriate order of magnitude constraints and demonstrate for the first time the viability of observing a DM induced drag on celestial objects.
It is worth noting that some studies call into question the need to impose a drag force to understand the orbit of G2, and claim a Keplerian fit is sufficient given the noisy data \cite{peissker2021apparent}. In the case of a gas cloud model, these conclusions would only strengthen the constraints on a DM induced drag force.

\section{Cross Section Projections}
\label{sec:cs_general}
From hereon we will focus on G2 since, as we have established above, it is a more sensitive probe of a dark drag force than the S-cluster stars. In particular, we show how G2's observations can be translated into constraints on DM–SM scattering cross sections across a wide range of DM masses. Throughout this analysis, we neglect general relativistic corrections. At the level of precision relevant here, the pericenter distance is large relative to Sgr.~A$^\star$'s Schwarzchild radius. 

We use the upper limit on the dissipative force on G2 to constrain the size of a DM-induced drag force, by mapping Eq.~\eqref{eq: a_D(r_p)} to Eq.~\eqref{eq: Fdrag eff}. This allows us to place a limit on the dark drag force: 
\begin{equation}
    |\vec{F}_{\rm drag }(r_p)| \lesssim M_{G2}\;|\vec{a}_D(r_p)|,
    \label{eq: drag comparison}
\end{equation}
where the drag force is effectively only acting during pericenter passage, following the arguments in Sec.~\ref{subsec: G2 orbit}. We can thus constrain the size of a dark matter interaction with the SM, for a given density at G2's most recent pericenter passing. While new information regarding the nature of G2 could alter these results, they nonetheless stand as a testament to the importance of modeling DM induced drag on galactic center objects.

\subsection{Light DM}
We begin by considering light DM, which in this context implies the drag force is largely independent of the DM mass. This occurs when the momentum transfer is proportional to $m_\chi$, canceling the $1/m_\chi$ dependence of the incident flux. In terms of mass, this corresponds to $ m_\chi \lesssim$ 1 GeV for nucleon scattering, and $m_\chi \lesssim$ 0.5 MeV for electron scattering. There are also lower DM mass limits in both cases, $m_\chi \gtrsim $ 1 eV, below which coherent enhancements may become relevant, as discussed in App.~\ref{sec: ultralight dm}. 

\begin{figure}[t]
    \centering
\includegraphics[width=\linewidth]{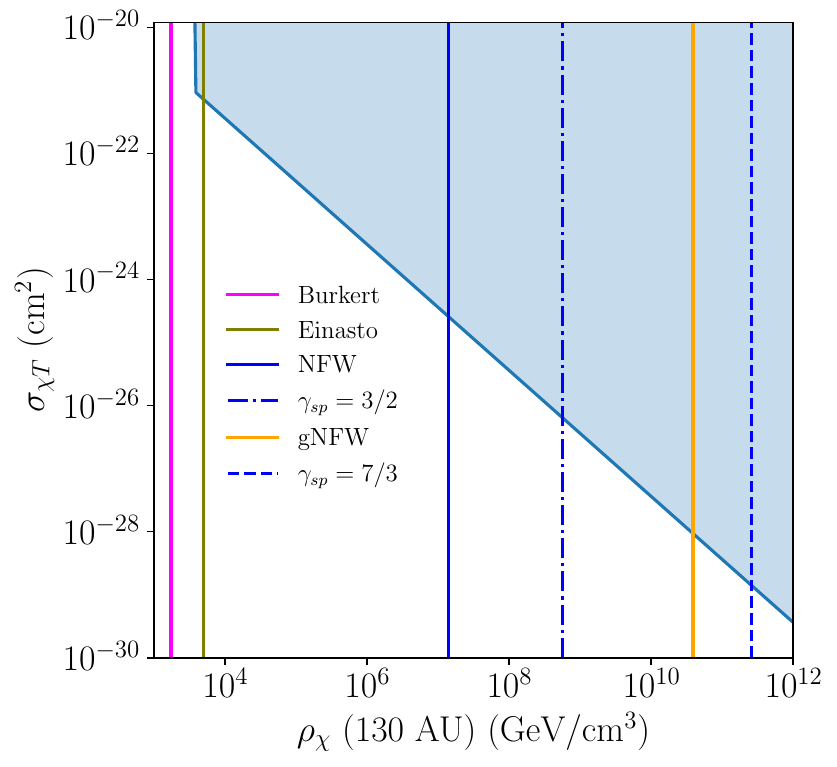}
        \caption{The blue shaded region corresponds to excluded DM-SM cross sections based on G2's orbital decay after its most recent pericenter passage. These lines apply to dark matter in the mass range $m_\chi \ll$ 1 GeV for nucleon scattering, and $m_\chi \ll$ 0.5 MeV for electron scattering. Vertical lines correspond to benchmark DM density profiles which give rise to the densities shown at G2's pericenter distance $r_p = 130 $ AU (see Fig. \ref{fig: density profiles}).}
    \label{fig: G2 density exclusion}
\end{figure}

We will consider, for simplicity, an isotropic scattering cross section with either electrons, $\sigma_{\chi e}$, or nucleons, $\sigma_{\chi n}$, under the assumption that G2's composition is entirely ionized hydrogen~\cite{gillessen2012gas, 2025arXiv251000897G}. 
Given the expected temperature of G2, $\mathcal{O}(10^4)$ K \cite{gillessen2012gas}, the thermal velocity of both electrons and protons is considerably smaller than that of G2's orbital velocity, so we assume either target is at rest. If the ionization fraction during pericenter passing is actually much less than unity, an atomic form factor must be included when considering electron scattering, but we leave this possibility to future work. The DM-target cross section that enters into Eq. \eqref{eq: Fdrag eff} (and consequently the constraint in Eq. \eqref{eq: drag comparison}) can thus be defined as 
\begin{equation}
    \sigma_{\chi T } = 
        \sigma_{\chi e} \text{ or } \sigma_{\chi n}.
\end{equation}

Figure~\ref{fig: G2 density exclusion} shows the inferred cross section limits from this procedure as a function of DM density at G2's latest pericenter distance. The vertical lines indicate for reference the density values predicted by the benchmark distributions from Sec.~\ref{sec:dm_GC_dist}. However, for sufficiently low DM densities, such as that predicted by cored distributions like the Burkert profile, no exclusion can be drawn. This loss of sensitivity is due to the fact that the maximum effective area in Eq.~\eqref{eq: Fdrag eff} saturates at the geometric cross section $\pi R_{\rm G2}^{ 2}$. As the sensitivity of the Einasto profile sits at large cross sections, and very close to the geometric limit, we do not include it in subsequent analyses. 

\begin{figure}[t]
    \centering
\includegraphics[width=\linewidth]{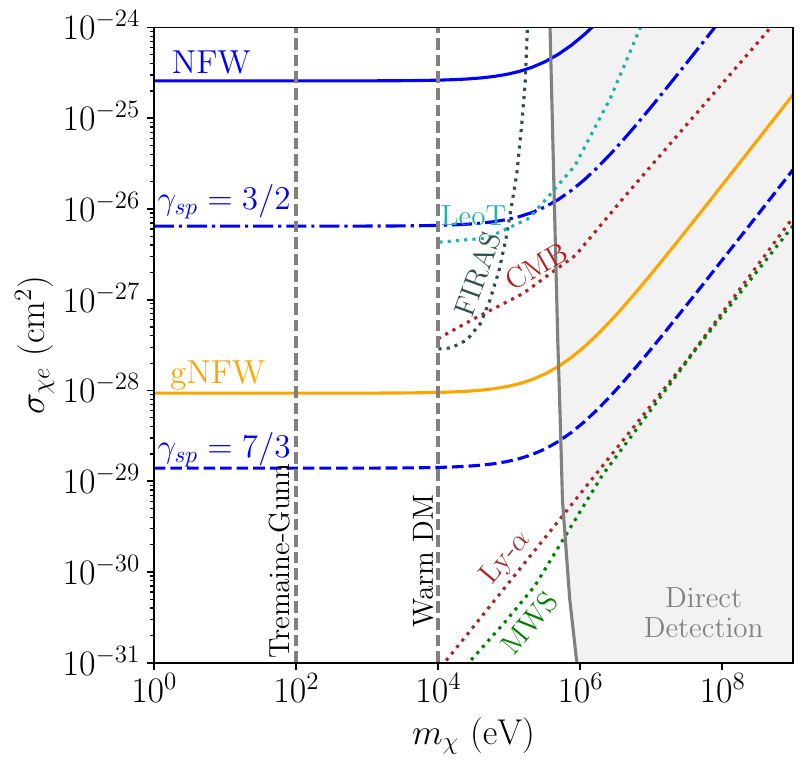}
        \caption{Excluded DM-electron cross section based on G2's orbital decay for various DM distributions as specified. Additional limits from FIRAS \cite{Ali-Haimoud:2015pwa}, CMB+BAO measurements \cite{Buen-Abad:2021mvc},
        LeoT cooling \cite{Wadekar:2019mpc}, and direct detection \cite{Hochberg:2021yud, Essig:2012yx, DarkSide:2022knj, SENSEI:2020dpa} are shown. The vertical dashed line approximately indicates the Tremaine-Gunn limit for light fermionic DM \cite{Tremaine:1979we}. Possible additional constraints are discussed in the text. 
        }
    \label{fig: electron sensitivity eV - GeV}
\end{figure}
\begin{figure}[t]
    \centering
\includegraphics[width=\linewidth]{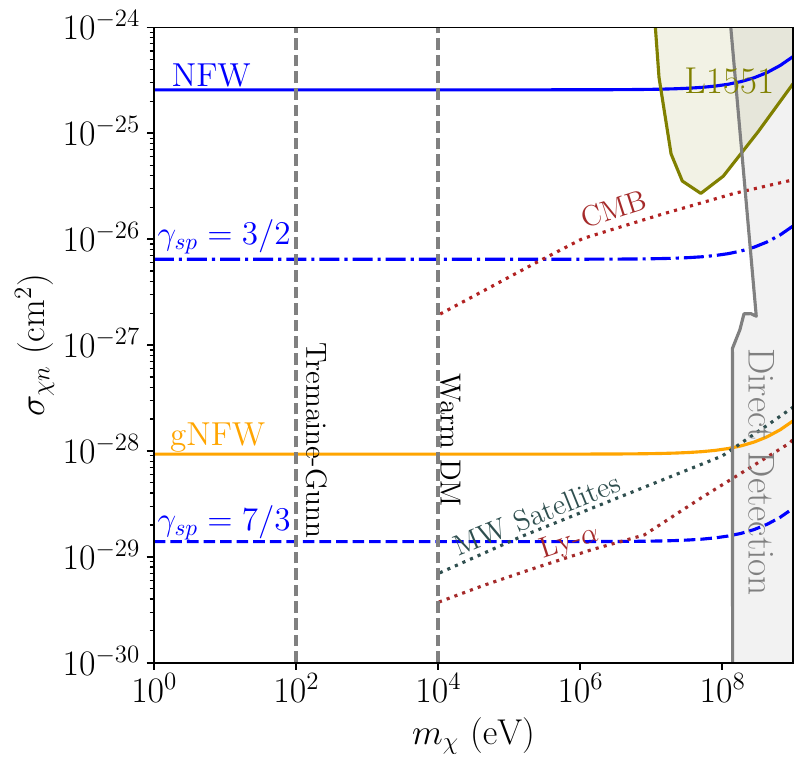}
        \caption{Same as Fig.~\ref{fig: electron sensitivity eV - GeV} for DM-nucleon cross section. Additional limits from direct detection (CRESST-Surface and XQC)\cite{Emken:2018run, Erickcek:2007jv}, L1551 gas cloud ionization \cite{Blanco:2023bgz} (see also \cite{Bhoonah:2018gjb, Bhoonah:2018wmw}), Milky Way satellites \cite{2021PhRvL.126i1101N}, as well as cosmological bounds derived from CMB \cite{Boddy:2022tyt, Gluscevic:2017ywp} and Lyman-$\alpha$ observations \cite{Rogers:2021byl}. Possible additional constraints are discussed in the text. 
        }
    \label{fig: nucleon sensitivity eV - GeV}
\end{figure}

In order to put these constraints in context, Figures~\ref{fig: electron sensitivity eV - GeV} and \ref{fig: nucleon sensitivity eV - GeV} respectively show electron and nucleon cross section sensitivity as a function of DM mass. In addition to the constraints derived from the DM density profiles listed in Fig.~\ref{fig: G2 density exclusion}, we show associated constraints from a variety of direct detection experiments in solid shaded regions, as well as cosmological constraints as dashed lines, with which our results are comparable in some regimes. 
Most cosmological bounds disappear for $m_\chi \lesssim 10$ keV because warm DM limits already constrain any thermal candidate below a few keV \cite{Viel:2013fqw}. We show all bounds associated with a velocity independent cross section for reference, but note that they can change vastly depending on the velocity dependence of interactions or for dark matter candidates with unique cosmological histories. For electron scattering, solar reflection also sets strong constraints in the mass region $1 \text{ keV}\lesssim m_\chi \lesssim 10 \text{ MeV}$ \cite{An:2017ojc,An:2021qdl,Emken:2021lgc}. However, as discussed in \cite{An:2021qdl}, for large enough cross sections, dark matter scatters too frequently off of the colder, outer regions of the sun, thus shutting off the enhancement received from kinetic up-scattering near the core. This transition occurs when $\sigma_{\chi e} \gtrsim 10^{-31} \text{ cm}^2$, and thus solar reflection limits are left off Fig.~\ref{fig: electron sensitivity eV - GeV}. For nucleon scattering, solar reflection does not enhance detection relative to virialized halo dark matter for $m_\chi \lesssim 0.1 \text{ GeV}$ \cite{Emken:2021lgc}.

Additional constraints on DM interactions arise from cosmic-ray up-scattering, which partially overlaps with the parameter space probed by G2’s orbital decay \cite{Bringmann:2018cvk,Dent:2019krz,Alvey:2019zaa,Cappiello:2018hsu,Cappiello:2019qsw,Bell:2021xff,Maity:2022exk,Bell:2023sdq}. However, because these processes involve relativistic energies, there is a non-trivial model dependency due to the specific cross-section scaling with the energy and masses of the intervening particles \cite{Dent:2019krz, Bardhan:2022bdg}. Such dependency may shift the inferred non-relativistic cross section limits towards lower values, leaving larger values unconstrained depending on the exact model realization. In contrast, the dark drag affecting G2’s orbit is inherently a non-relativistic effect. Albeit weaker, our constraints are more robust in terms of the DM-SM interaction, as they do not require extrapolation from the high-energy regime. Note that there are also constraints from blazar-boosting at these masses \cite{Granelli:2022ysi}, though we do not consider them explicitly as these become relevant at much lower cross-sections and the DM content around these objects is more uncertain. 

Finally, it is worth mentioning that one needs to be careful when constructing a dark matter model with the cross sections shown in Figs. \ref{fig: electron sensitivity eV - GeV} and \ref{fig: nucleon sensitivity eV - GeV}. Strong limits exist on generic models coming from the combination of stellar cooling, supernova cooling, and Kaon decays \cite{Elor:2021swj}. Such constraints can be avoided if DM is composite, for example, but we do not discuss explicit constructions here as the purpose of this work is to demonstrate the potential of using kinematic observations to detect dark matter, rather than rule out particular models.

\subsection{Heavy Inelastic DM}
\label{sec:heavy_inel_DM}
We now proceed with our scan over the parameter space probed by G2's trajectory with DM above the GeV-scale. In this case, both spin-independent and spin-dependent elastic scattering cross-sections are tightly constrained by direct detection experiments \cite{XENON:2019rxp,PandaX-4T:2021bab,DarkSide-50:2022qzh,LZ:2022lsv}, reaching values far below the electroweak scale, and therefore well beyond what can be probed with the observations of G2's orbital decay. A notable exception, however, arises for inelastic dark matter, where scattering involves a transition between states of different masses, $i.e.$ $\chi_{1(2)} + {\rm SM} \rightarrow \chi_{2(1)} + {\rm SM}$ with a mass splitting 
\begin{equation}
    \delta m_\chi = m_{\chi_2} - m_{\chi_1}~.
\end{equation}
For interstate mass splittings beyond a few hundreds of keV, endothermic scattering, where halo dark matter up-scatters into a heavier state inside a terrestrial detector, is kinematically suppressed, since the typical energies become insufficient for the transition to be excited, even in the limit that $m_\chi \gg m_N$. For models in which there is no sizable relic component $\chi_2$ today, exothermic scattering (which requires no minimum energy) cannot occur. In this case, direct detection constraints rapidly disappear with increasing mass splitting, as the rate relies ever more on the high-velocity tail of the halo distribution. This inelastic scenario is often invoked to reconcile null direct searches with astrophysical excesses that could be associated with DM, see $e.g.$ \cite{Tucker-Smith:2001myb,Chang:2008gd,Finkbeiner:2007kk,Pospelov:2007xh,Batell:2009vb,Zhang:2016dck,Alvarez:2019nwt,Hooper:2025fda}. Currently, CRESST-II sets the strongest constraints on inelastic cross-sections, at the level of $\sigma_{\chi n} \lesssim  10^{-33} \ \rm cm^2$ for $\lesssim 400 \ \rm keV$ splittings and DM masses $m_\chi \gtrsim 1 \ \rm TeV$, owing to their use of heavy tungsten W$^{180}$ as the target \cite{Bramante:2016rdh}. For splittings between $400 \ {\rm keV} \lesssim \delta m_\chi \lesssim 1.2 \ \rm MeV$, recent searches relying on the induced transition of nuclear states extend the cross-section limits beyond what CRESST-II can probe \cite{Lehnert:2019tuw,Alves:2023kek}, albeit at larger cross-sections.

G2's large pericenter velocity makes it a viable probe to study inelastic DM scattering processes, as the kinematic threshold for endothermic scattering can be met for larger mass splittings compared to direct detection. In the non-relativistic limit, the maximum splitting that can kinematically be excited is given by
\begin{equation}
    \delta m_\chi^{\rm max} = \frac{1}{2} \, \mu_{\chi N} \, v^2 \simeq 745 \ {\rm keV} \, \left(\frac{m_N}{3.72 \ \rm GeV}\right) \left(\frac{v}{0.02}\right)^2~,
    \label{eq:deltamchi_max}
\end{equation}
where $\mu_{\chi N}$ is the reduced mass of the DM-nucleus system and, in the rightmost expression, we have normalized it to the case of DM$-^{4}$He scattering in the limit $m_\chi \gg m_N$ as G2 passes through pericenter. Each element in G2's composition thus probes a certain mass splitting range, and so the drag force must technically be computed by integrating over the elemental composition of G2. It is clear from Eq.~\eqref{eq:deltamchi_max} that heavier elements in G2 probe larger mass splittings. However, we will primarily focus on coherent scattering against helium nuclei. As we detail below, while including heavier elements in this analysis yields stronger reach in mass splitting, the resulting limits also become considerably more model-dependent. For simplicity, we will assume G2's helium abundance matches the solar abundance \cite{2009ARA&A..47..481A}, roughly $10^{-1.07}$ relative to hydrogen by number. This choice should be conservative, as a higher abundance would boost the sensitivity, and metallicity measurements of the Galactic Center indicate values exceeding those of the Sun \cite{2014arXiv1409.2515R,2017MNRAS.464..194F}.

For this inelastic scenario, we approximate the full DM-target cross-section as (see $e.g.$ \cite{Garani:2017jcj,Busoni:2017mhe,Batell:2009vb}) 
\begin{equation}
   \sigma_{\chi T } = \sigma_{\chi n}\, A^4 \left(\frac{m_\chi+m_n}{m_{\chi}+A m_n}\right)^2 |F(E_R)|^2 \, \sqrt{1 - \frac{\delta m_\chi}{\delta m_\chi^{\rm max}}}~,
    \label{eq:cross_section_nucleus}
\end{equation}
where $\sigma_{\chi n}$ is the DM-nucleon cross-section, $A$ is the mass number of scattered nucleus, and $\delta m^{\rm max}_\chi$ is the maximum mass splitting allowed by kinematics. The function $|F(E_R)|^2 = \exp\left(-{E_R}/{E_N}\right)$ is the Helm form factor evaluated at the recoil energy $E_R$ \cite{Helm:1956zz}, which accounts for nuclear structure effects at sufficiently large energy exchanges. The scale $E_N = 3/(2m_N\Lambda_N^2)$ is determined by the nuclear radius $\Lambda_N \simeq \left(0.91 \left({m_N}/{\rm GeV}\right)^{1/3}+0.3\right) \, \rm fm~$. 

For inelastic scattering, the minimum recoil energy is no longer zero,   
and both $E_R^{\rm min}$ and $E_R^{\rm max}$ become functions of the interstate splitting. The recoil range now spans \cite{Shu:2010ta}
\begin{equation}
    E_R = m_N v^2 \left(1 - \cos(\theta_c) \sqrt{1 - \frac{\delta m_\chi}{\delta m_\chi^{\rm max}}}\right) - \delta m_\chi~,
    \label{eq:recoil_range_inel}
\end{equation}
with $\theta_c$ ranging from 0 to $\pi$, corresponding to $E_R^{\rm min}$ and $E_R^{\rm max}$ respectively. For brevity, we have written Eq.~\eqref{eq:recoil_range_inel} in the limit of $m_\chi\gg m_N$, but in our analysis we use the full expression as shown in \cite{Shu:2010ta}. In the limit that $\delta m_\chi \rightarrow 0$, Eq.~\eqref{eq:recoil_range_inel} reduces to the recoil energy of an elastic scattering event as expected.  

Relative to the elastic, light DM regime previously analyzed, the loss of coherence at large momentum exchanges introduces an anisotropy in the scattering angle. Simply put, momentum exchanges with a large projection onto G2's direction of motion are now suppressed through the Helm form factor. To incorporate the effects of the Helm form factor in nuclear scattering without integrating over the full recoil spectrum, we approximate it at an average recoil energy $|F(E_R)|^2 \rightarrow |F(\langle E_R \rangle)|^2$ defined by 
\begin{equation}
    \langle E_R \rangle = \frac{\int_{E_R^{\rm min}}^{E_R^{\rm max}} E_R \, |F(E_R)|^2 \, dE_R}{\int_{E_R^{\rm min}}^{E_R^{\rm max}}  |F(E_R)|^2 \, dE_R}~.
    \label{eq:average_recoil}
\end{equation}
We justify this approximation and parameterize the modification to the drag force by introducing a function $f(\theta^*_c)$ in Appendix~\ref{App: f theta_c}. This lets us write the drag force in a manner similar to Eq. \eqref{eq: Fdrag eff} given by 
\begin{equation}
\label{eq:drag_with_theta_c}
  |\vec{F}_{\rm drag}| =  f(\theta^*_c) \, \left(\frac{m_{T}}{m_\chi}\right)\rho_\chi A_{\rm eff}(\theta^*_c) v^2.
\end{equation}
The function $f(\theta^*_c) \in (0,1)$ indicates the penalty to $\langle\vec{q}\rangle$ in 
Eq. \eqref{eq: q avg}, with $\theta^*_c$ the center of mass scattering angle associated with the average recoil energy in Eq. \eqref{eq:average_recoil}. The effective area $A_{\rm eff}$, introduced in Eq.~\eqref{eq: A_eff}, now also depends on the angle $\theta^*_c$ because of the energy dependent cross section. 

\begin{figure}[t]
    \centering
\includegraphics[width=\linewidth]{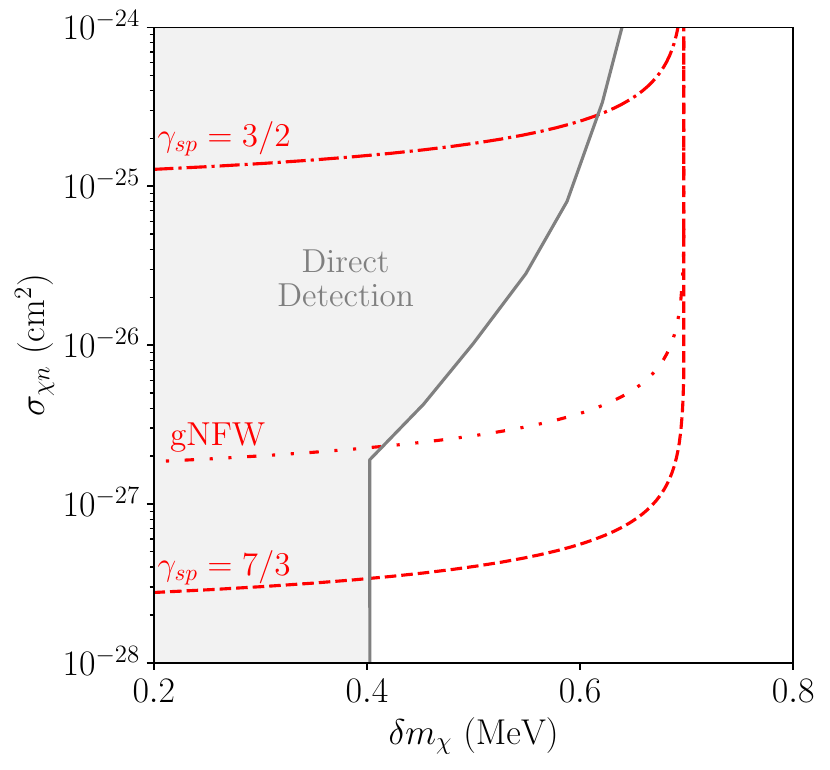}
    \caption{Excluded inelastic DM-nucleon cross section based on G2's orbital decay for various distributions as a function of mass splitting, for $m_\chi = 1$ TeV, due to scattering with Helium. Dot-dashed lines are set by a relaxed spike profile, dotted lines are set by a gNFW profile, and dashed lines are set by a full spike profile. We also include limits from direct detection shaded in gray \cite{Lehnert:2019tuw, Bramante:2016rdh}.}
    \label{fig: sigma sensitivity inelastic}
\end{figure}

Figure~\ref{fig: sigma sensitivity inelastic} presents the inelastic cross section constraints as a function of mass splitting, incorporating the corrections to the drag force due to inelasticity and the Helm form factor. For concreteness, we have fixed the mass to $m_\chi = 1 \ \rm TeV$; the mass dependence of the results is addressed below. As anticipated, we see that G2's orbital decay covers mass splittings currently inaccessible to underground experiments, reaching splittings up to $\sim 750 \ \rm keV$. Overall, the small nuclear radius of helium implies small form factor suppression, and $f(\theta^*_c) \simeq 1$. As a result, the lines remain relatively flat except when close to $\delta m_\chi^{\rm max}$, in which case the scattering phase space becomes increasingly restricted. We emphasize that these results are conservative in that we have likely underestimated the abundance of helium by taking solar-level values. 

The DM mass dependence of these results, in the limit $m_\chi \gg m_N$, is determined by the DM number density $\rho_\chi / m_\chi$, the Helm form factor, and the reduction in (parallel) momentum transfer encapsulated in $f(\theta^*_c)$. Of these, the flux factor enhances cross-section sensitivity as $1/m_\chi$, while the latter two depend on the DM mass through the average recoil energy $\langle  E_R \rangle$, $cf.$ Eq.~\eqref{eq:average_recoil}. For masses well above the target mass, both the form factor suppression and the momentum transfer penalty become nearly independent of the DM mass, as $\langle E_R \rangle$ simply scales with nucleus mass in this limit. If the mass is instead closer to the target mass, momentum transfer becomes slightly more efficient, but the maximum mass splitting is reduced. Therefore at lower masses, the sensitivity is enhanced away from the maximum splitting, but the form factor suppression becomes relevant at smaller mass splittings. In summary, the limits for $m_\chi = 1 \ \rm TeV$ shown in Fig.~\ref{fig: sigma sensitivity inelastic} retain the same qualitative features with varying mass, but with a normalization that scales approximately as $1/m_\chi$ due to the number density dependence and a slight variation in the range of probed splittings. 

Finally, we comment on DM scattering off heavier elements, as well as the related model dependence of our constraints. For scattering off heavier elements, the parameterization of the cross-section as given by Eq.~$\eqref{eq:cross_section_nucleus}$ combined with Eq.~\eqref{eq:average_recoil} is no longer a reasonable approximation, due to the sharp variation of the Helm form factor across the kinematic range. In this case, the drag force must be computed from integrating the full differential cross-section (see App. \ref{App: f theta_c}).  
While this could probe larger mass splittings, an appreciable form factor suppression in the cross section combined with potentially lower abundances generally lead to weaker constraints than Helium, despite the larger atomic number.
Moreover, as noted in \cite{Digman:2019wdm}, the scaling of Eq.~\eqref{eq:cross_section_nucleus} with mass number can become unreliable at sufficiently large cross-sections, due to the breakdown of the Born approximation used in its derivation. For contact interactions, this implies that higher order partial waves must be included in the computation of the cross-section, which in the repulsive case at most saturates to a few times the geometric cross-section of the target nucleus. Attractive contact interactions present a more complex scenario, as they can support the formation of bound states, and the associated resonances can drive the cross-section above the nucleus' geometric size. In either case, the inclusion of higher partial waves leads to a scaling that generically deviates from $A^4$ for strong enough couplings of the underlying model (note, however, composite DM models with cross-sections $\propto A^4$ can in principle be constructed \cite{Acevedo:2024lyr}).


For DM-helium scattering, however, the DM-nucleon cross-sections shown in Fig.~\ref{fig: sigma sensitivity inelastic} translate into nuclear-level cross-sections at or near the regime where the scaling relation used in Eq.~\eqref{eq:cross_section_nucleus}  is generically valid. For nucleon cross sections  $\sigma_{\chi n} \gtrsim 10^{-27} \text{ cm}^2$, the results become slightly model dependent, but Eq. \eqref{eq:cross_section_nucleus} remains a good approximation of competing model assumptions (see \cite{Digman:2019wdm} for more details). 

\section{Dark Drag and the Missing Giants}
\label{sec:miss_giants}
 Having examined in the previous sections the effects of a dark drag force on a single orbital timescale, we now turn to its long-term impact on the population of objects around Sgr.~A$^\star$. Specifically, we focus on the missing giants problem: a well-established under-abundance of red giant and horizontal branch stars in the innermost region of the Milky Way's nuclear star cluster. 
Basic relaxation arguments indicate that these stars, being among the most evolved within the cluster, should follow a cuspy distribution \cite{1976ApJ...209..214B, 2006ApJ...645L.133H, Alexander:2008tq}. Instead, red giant counts in this region exhibit either a flattening or inversion of its surface density toward projected radii $ r\lesssim 0.1 \ \rm pc$, which has come to be known as the ``missing giants problem" \cite{Schodel:2007er, 2009ApJ...703.1323D, 2009A&A...499..483B, 2010ApJ...708..834B, 2018A&A...609A..26G, 2019ApJ...872L..15H}. Several mechanisms have been proposed to explain this paucity. These include stellar collisions between giants and main-sequence stars \cite{1996ApJ...472..153G, 2009MNRAS.393.1016D, 2023ApJ...955...30R}, interactions with accreted matter and jets of Sgr.~A$^{\star}$ \cite{2025MNRAS.540.1586K}, selective envelope stripping of the star-forming disk \cite{2014ApJ...781L..18A}, or even ejections by a recent merger \cite{2010ApJ...718..739M}. However, these channels are individually unable to remove red giants at the level that is required to explain the paucity of these objects toward the center. 

We show here that for currently unconstrained parameter space, the dark drag-induced orbital decay of red clump and horizontal branch stars could result in their tidal disruption by Sgr.~A$^{\star}$, given their large characteristic sizes, and those of their giant branch predecessors, of the order $\sim (10-100) R_\odot$ . This naturally constitutes a mechanism that could sizably remove the giant branch stars within the region of observed depletion. However, a realistic expectation might be that the dark drag acts in conjunction with the aforementioned phenomena. For instance, stellar collisions between main-sequence and red giant stars are expected to be relatively frequent at the projected radii where the latter population is depleted. 

Ref.~\cite{2018A&A...609A..26G} analyzed the distribution of red clump stars of magnitudes in the $K$-bands ranging $12-16$ and $> 18$, and found a flattening of the distribution at radii $\lesssim 1 \ \rm pc$. More recently, \cite{2019ApJ...872L..15H} spectroscopically identified a density cusp of late-type stars down to $r \gtrsim 0.025 \ \rm pc$. We focus on these scales and show that, within a region of this size, stars can be efficiently tidally-disrupted from the dark drag-induced orbital decay.

To this end, we semi-analytically estimate the orbital decay of a star, starting from a given semi-major axis $a_0$ and eccentricity $e_0$, by integrating over the energy and eccentricity loss during each pericenter passage, motivated by the procedure of \cite{Sz_lgy_n_2022}. For a given $a_0$ and $e_0$, we start with the initial specific orbital energy $\varepsilon_0$ and specific angular momentum $h_0$,
\begin{equation}
\label{eq: E0}
    \varepsilon_0 = -\frac{G \, M_{\rm SgA^{*}}}{2a_0}~,
\end{equation}
\begin{equation}
\label{eq: h0}
    h_0 = \sqrt{G \, M_{\rm SgA^{*}} \, a_0 (1-e_0^2)}~,
\end{equation}
where we fix the center of mass to the position of Sgr.~A$^{\star}$. The dissipative drag force results in changes to $\varepsilon$ and $h$ on each orbit given by 
\begin{equation}
\label{eq: delta E_orb}
\Delta \varepsilon = \left(\frac{\vec{v}\cdot \vec{F}_{\rm drag}}{M_\star}\right)\Delta t~,
\end{equation}
\begin{equation}
\label{eq: delta h_orb}
\Delta h = \left(\frac{|\vec{R}\times \vec{F}_{\rm drag}|}{M_\star}\right)\Delta t~,
\end{equation}
where $\vec{R}$ and $M_\star$ are the position and mass of the orbitally-decaying star, $\Delta t$ is the characteristic timescale of the pericenter passage, and we take $\vec{F}_{\rm drag}$ as given by the geometric limit, i.e. Eq. \eqref{eq: Fdrag simple}. To compute these perturbative changes, we evaluate $\vec{F}_{\rm drag}$ at pericenter, and make the approximation that $\Delta t = D_p/v_p$ where
\begin{equation}
    \label{eq: vp}
    v_p = \sqrt{\frac{GM_{\rm SgA^*}(1+e)}{a(1-e)}}~,
\end{equation}
is the velocity at pericenter, and 
\begin{equation}
\label{eq: Dp}
    D_p = (2-e)\pi r_p,
\end{equation}is the approximate distance traveled during the pericenter phase ($i.e.$ a near semi-circle distance with radius equal to the pericenter distance $r_p = a(1-e)$). The factor of $(2-e)$ in $D_p$ interpolates between high eccentricities where a semi-circle of radius $r_p$ is more appropriate, and low eccenricities where a full circle diameter is more appropriate. We only consider changes to energy and eccentricity during pericenter phase, up until the point where the eccentricity is low enough that they are effectively equal (chosen to be $e < 0.05$). Past this point, we ignore $\Delta e$ and compute $\Delta \varepsilon$ by taking $D_p = 2\pi r_p$. Before this point, however, we update the eccentricity during each orbit according to 
\begin{equation}
    \label{eq: e update}
    e_{i+1} \simeq e_i + \Delta e = \sqrt{1+\frac{2(\varepsilon_i + \Delta \varepsilon)(h_i + \Delta h)^2}{(GM_{\rm SgA^{*}})^2}}~,
\end{equation}
where the subscript $i$ denotes the current iteration step. This perturbative approximation requires that the changes $\Delta \varepsilon$ and $\Delta h$ are small compared to $\varepsilon_i$ and $h_i$, a requirement that always holds in our analyses given that the gravitational force of Sgr.~A$^\star$ significantly dominates over the drag force. With $\Delta \varepsilon$ and $\Delta e$ at hand, we can numerically calculate the time it takes a star to fall from its initial orbit to its tidal radius by tracking the orbital time associated with each update until $r_p \leq R_{\rm tid}$, which we approximate as 
\begin{align}
\label{eq: R_tidal}
    R_{\rm tid} &\simeq R_* \left( \frac{M_*}{M_{\rm SgA^{*}}} \right)^{-1/3} \nonumber \\
    &\simeq 37 \, {\rm AU} 
    \left( \frac{R_*}{50 \, R_\odot} \right)
    \left( \frac{M_*}{M_\odot} \right)^{-\frac{1}{3}}
    \left( \frac{M_{\rm SgA^{*}}}{4 \times 10^6 \, M_\odot} \right)^{\frac{1}{3}}~.
\end{align}

\begin{figure}[t]
    \centering
\includegraphics[width=\linewidth]{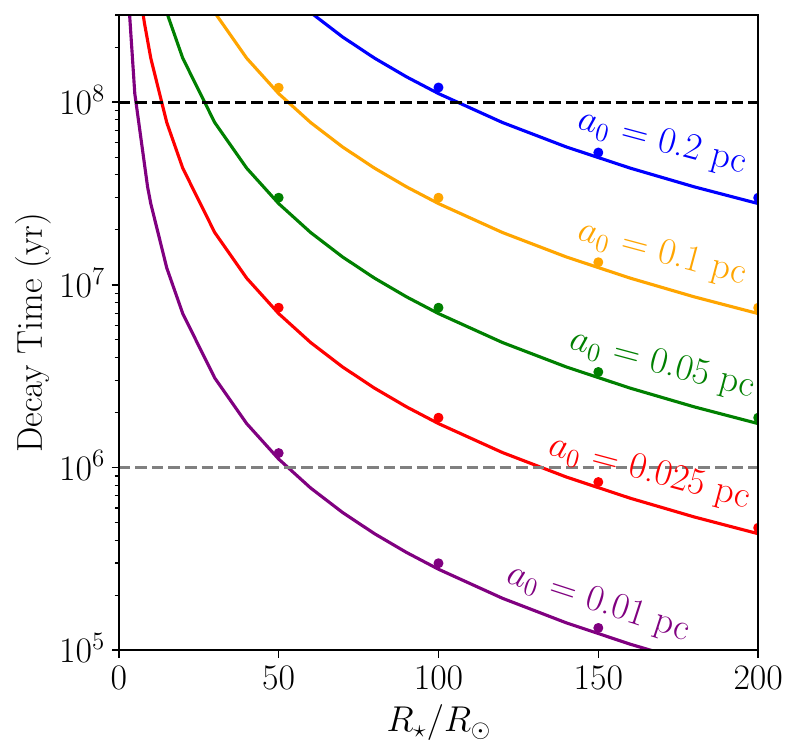}
        \caption{Timescale required for a star to be tidally disrupted, obtained from semi-analytic calculations starting from various initial orbits with semi-major axes $a_0$ as specified. In all cases, we assume a gNFW density profile, a stellar mass $M_\star = M_\odot$, and an initial eccentricity given by the relaxed cluster average $e_0  = \langle e \rangle = 0.66$. Black and gray dashed lines correspond to the typical lifetime of a red clump star and the typical age of early-type stars (e.g. S2) near the galactic center, respectively. 
        The circles show the decay timescale obtained from simulations.}
    \label{fig: decay times}
\end{figure}

Figure~\ref{fig: decay times} shows the resulting decay times for stellar mass sized objects, which can be considerably shorter than the typical lifetime of stars in the red giant branch, of order $0.1 \ \rm Gyr$ \cite{2016ARA&A..54...95G}. Isotropically distributed and relaxed stars near the galactic center evolve in orbits that are effectively Keplerian about Sgr.~A$^\star$, and follow a normalized eccentricity distribution of $n(e) = 2e$, see $e.g.$ \cite{Alexander:2005jz}. For concreteness, we fix initial eccentricities to the average value of $\langle e \rangle=2/3$ in Fig.~\ref{fig: decay times}. To confirm the results from the analytical approach above, we have additionally simulated the dark drag-induced orbital decay starting from the same initial conditions utilizing a Velocity Verlet algorithm \cite{hairer2006geometric}, for a few example points in parameter space. 

We find excellent overall agreement between our theoretically computed orbital decay times and those obtained from the simulation. Despite only computing energy loss at pericenter and using a simple estimate of pericenter distance $D_p$, this agreement occurs because of the increase in DM density towards lower galactic radii and the scaling of the drag force with $v^2$. While we do not show the eccentricity evolution explicitly, we find a discrepancy between the eccentricity decay rate relative to the simulation of at most a factor of two.
Including energy and eccentricity updates during both pericenter and apocenter phases of the orbit, as is done in \cite{Sz_lgy_n_2022}, may correct for this difference as energy loss during apocenter tends to increase eccentricity, while energy loss during pericenter decreases it. However, as our main goal is to estimate the total orbital decay time, we find our approximation sufficient (see Fig. \ref{fig: decay times}).  Note that in both analyses, we did not include the effects of dynamical friction, as we find these negligible compared to the DM drag. This is because we are considering large-scale celestial objects for which hydrodynamic drag forces are potentially very significant, whereas dynamical friction by contrast tends to dominate in compact ones (see $e.g.$ Ref.~\cite{2015ApJ...811...54G}). 

\begin{figure}[t]
    \centering
\includegraphics[width=\linewidth]{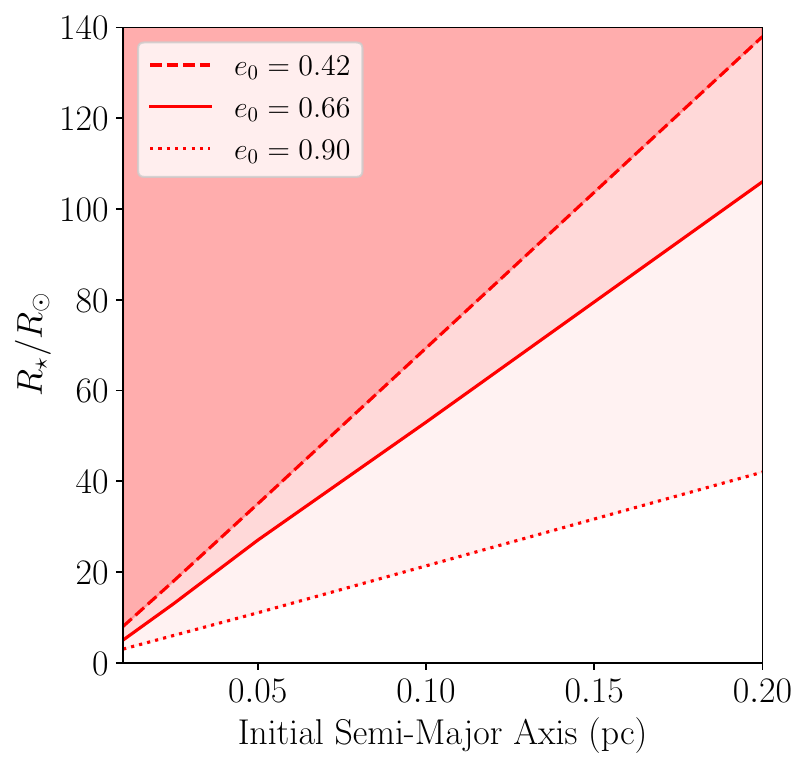}
        \caption{Parameters for which a star's orbital decay time is less than the typical lifetime for a red giant, $t\simeq 0.1$ Gyr. The shaded region corresponds to stars that would have fallen into Sgr.~A$^\star$ due to dark drag forces, given a gNFW dark matter density profile, a stellar mass $M_\star = M_\odot$, and an initial eccentricity $e_0 = \langle e \rangle \pm \sigma_e$, assuming geometric scattering.}
    \label{fig: missing giant params}
\end{figure}

Figure~\ref{fig: missing giant params} explicitly shows the stellar radii for which a solar-mass star would have reached its tidal radius within the typical lifetime of a red giant ($i.e.$ has a decay time less than 0.1 Gyr) as a function of initial semi-major axis. In this plot, we include eccentricities within one standard deviation $\sigma_e$ of the expectation value $\langle{e}\rangle$ (as given by the relaxed eccentricity distribution above), in order to illustrate the variation of the decay times with this quantity. As expected, higher initial eccentricities result in much shorter orbital decay timescales. This is because of the smaller pericenter distance, where the DM density is higher, combined with higher pericenter velocities, both of which lead to a significantly larger drag force in each passage relative to lower eccentricity values. Consequently, even stars with $R_\star \ll 100 R_\odot$ can experience rapid orbital decays if on eccentric enough orbits. Moving forward, it will be interesting to analyze whether more stringent constraints on DM$-$SM portals can be obtained solely based on the survival times of observed stars, especially those with high eccentricity for which a substantial sample is available. However, such an analysis would require detailed knowledge about the initial conditions and the star formation history of the Milky Way's nuclear star cluster. Additional modeling of star-star interactions in the S-cluster as well as the back reaction of stellar interactions with DM onto the DM distribution would also be required for such a study. We leave this for future work. 

Finally, we emphasize that the cross-sections ruled out by our conservative analysis of G2's orbital decay are not required to obtain such short decay times for late-type stars. To approximate the cross section required to saturate the geometric limit of a red giant star, we require the mean free path of a dark matter particle passing through the outer most regions of the star to be less than the radius of the star. The geometric limit will therefore be roughly saturated when

\begin{equation}
    \sigma_{\chi T} \sim \frac{m_n}{\rho( R_\star)\times R_\star }.
\end{equation}
Practically speaking, we cannot use the very edge of the photosphere, as the distance the dark matter particle travels through the star will be negligible. Taking instead the typical density slightly below the surface of $10^{-7}$ g/cm$^2$ \cite{Ryu2024}, one finds a cross section of $\sigma_{\chi T}\approx 10^{-31} \text{cm}^2$ for $R_\star = 50R_\odot$. This value really captures the geometric limit of an area somewhat below the surface, but gives a good approximation nonetheless.
However, a drag force as strong as Eq.~\eqref{eq: Fdrag simple} is not required to obtain orbital decay times comparable to its lifetime. This can be seen by direct comparison with Figs.~\ref{fig: decay times} and \ref{fig: missing giant params}, which show decay times assuming the geometric limit. As the drag force has an approximately linear dependence with cross-section, one can see that for sufficiently small initial semi-major axis and/or sufficiently large initial eccentricity, much lower cross sections than the geometric value above will achieve decay times $\lesssim 0.1 \ \rm Gyr$. Indeed, for stars with $a_0 \sim 10^{-2} \ \rm pc$ and radii $\gtrsim 100 R_\odot$, tidal disruption may occur before reaching the end of its lifetime for cross sections as low as $10^{-34} - 10^{-35} \ \rm cm^2$. 
Such values are largely unconstrained for sub-GeV DM with nucleon portals or inelastic DM with $\gtrsim 500 \ \rm keV$ mass splittings. Alternatively, one can also consider shallower density profiles rather than smaller cross sections.

\section{Summary and Outlook}
\label{sec:conclusion}
We have analyzed the dynamical effects that dark matter may have on orbiting objects in the Galactic Center when a Standard Model portal is present. Dark matter interactions with either nucleons or electrons induce a dissipative drag force through frequent momentum exchanges as the stars orbit around Sgr.~A$^\star$ and its surrounding dark halo. The high dark matter density of this central region, combined with the large pericenter velocities of stars, results in a drag force that can be significant, potentially leading to orbital decays on timescales much shorter than the age of the Milky Way's nuclear star cluster and, in some cases, the lifetime of the stars themselves.

We first studied the short timescale changes induced by a dark drag force on objects orbiting around the Sgr.~A$^\star$ with low pericenter distances. In particular, we focused on G2, a large gas cloud observed to orbit Sgr.~A$^\star$ with well-determined orbital parameters. Utilizing existing data which set complementary limits on drag forces from Sgr.~A$^\star$'s baryonic accretion, we set constraints on the strength of the dark matter-Standard Model portal for both electron and nucleon cross-sections, across a wide range of masses, spanning from the eV- to the TeV-scale. Our constraints naturally depend on the halo distribution around Sgr.~A$^\star$, however, we find that even for conservative choices that do not assume a DM spike, the orbital decay limit of G2 can complement other probes; particularly for sub-GeV masses where direct detection experiments are limited by minimum recoil energies. We have also found observations of G2 place strong constraints on heavy inelastic dark matter, due to its high velocity during pericenter passage. Under simplified assumptions on the composition of G2, we have shown how its orbital decay extends inelastic cross-section existing limits into the nearly MeV mass splitting regime. While we focused on helium scattering, other heavier elements can in principle extend these constraints, although the inferred cross-section bounds are typically more model-dependent. We also considered additional nearby stars, which potentially can provide substantially more cross section sensitivity than G2. While these objects have lower astrophysical uncertainties in terms of pericenter behavior, an observable signal would require increased dark matter densities in tension with limits on the extended mass around Sgr.~A$^\star$ \cite{2020A&A...636L...5G}.

We have also explored the long-term impact of this drag force on the population of late-type giant stars. Intriguingly, for currently unconstrained interaction strengths, the dark drag could help explain the observed apparent depletion of red giant and horizontal branch stars in the innermost region of the Galactic Center. We explored this possibility both semi-analytically and through numerical simulations of stellar orbital decay under dark drag. By scanning over initial orbital elements and stellar properties, we have found that a dark drag force could induce sufficient orbital decay to cause the tidal disruptions of large-size stars by Sgr.~A$^\star$, on timescales much shorter than their typical lifetime. This is the case even for more moderate dark matter profiles that do not assume a spike is formed at any stage of the Milky Way's evolution. 

More broadly, we have shown that the dark hydrodynamic drag resulting from dark matter interactions with the visible sector could produce long-term secular changes to the population of stars in systems where the dark matter density is large, even if it does not dominate the gravitational potential. This opens a number of additional directions that could be explored. For example, we expect the standard picture of mass segregation \cite{Alexander:2008tq} $-$the equipartition effect whereby upon relaxation heavier stars sink toward center of the potential well$-$ to be modified in dark matter-dense systems, as now light objects with large radii potentially sink along with the heavier population into the potential well. This could alter the observed distribution of stars in old, dynamically-relaxed systems such as globular clusters, the Galactic Center, and dwarf spheroidal galaxies. On smaller scales, the dark drag could also accelerate the in-spiral of binary systems in which a main-sequence or red giant star orbits a white dwarf, a possible progenitor of Type-Ia supernovae \cite{Han:2003uj,Wang:2018pac}. Another interesting venue would be the evolution of Population III stars, which likely formed early on in regions of high dark matter density. The dark drag effect could increase the rate of tidal disruption events by black holes, a future detection prospect for Population III stars with both the James Webb and Roman telescopes \cite{Chowdhury:2024wvm} and gravitational wave observatories \cite{2025arXiv250418615P}, as well as a potential component of the formation and growth of supermassive black holes \cite{Wang:2025zzg}. In a related manner, it will also be interesting to examine the potential role of a dark drag force in galaxy assembly, considering James Webb observations favoring it at redshifts larger than expected from the cold dark matter paradigm \cite{2023ApJ...955...13B,2023MNRAS.519.1201A,2024ApJ...960...56H}. We drag along these questions for future work. \\

\textit{Note added:} Shortly after this work was announced on arxiv, ref.~\cite{Gustafson:2025ypo} appeared. Their analysis concerning S2's evolution is in agreement with the simple estimates discussed in Sect.~\ref{sec:drag_targets}. As noted here, these constraints primarily require dark matter densities considerably larger than those of the profiles we consider. 



\acknowledgements 
We thank Stefan Gillesen, Rebecca Leane, Dave McKeen, Adam Ritz, Natalia Toro and Weishuang Linda Xu for helpful comments and discussions. JFA, AJR and LSO are supported by the U.S. Department of Energy under Contract DE-AC02-76SF00515. AJR is also supported by the NSF GRFP under grant DGE-2146755. This research was also undertaken thanks in part to funding from the Natural Sciences and Engineering Research Council of Canada through the Arthur B. McDonald Canadian Astroparticle Physics Research Institute.

\clearpage
\newpage
\onecolumngrid
\appendix
\section{Free Molecular Flow Limit}
\label{app: Drag Force}
In this section we will discuss our assumption that we can always work in the free molecular flow limit. This limit applies when the mean free path of the gas that an object moves through is larger than the size of the object. In this way, self-interactions of the gas can be ignored when deriving a drag force, as a particle scattering off the object will not disrupt the surrounding distribution before it leaves the vicinity of the object. The force can therefore be derived considering only the momentum that a particle in an initially thermal gas imparts when scattering off the object \cite{Baines:1965}. In the case of dark matter, self-interactions are typically assumed to be small, and this limit is therefore valid. However, as we will at times consider the drag induced by very high dark matter number densities, it merits understanding when this assumption breaks down. A more precise statement of the free molecular flow limit requires that the Knudsen number be much larger than 1 \cite{TAGHAVINEJAD2020103535}, $i.e. $
\begin{equation}
    K_n = \frac{\lambda_\chi}{R_\star} \gtrsim 10~,
\end{equation}
where $R_\star$ is the radius of the object in question, and $\lambda_\chi$ is the mean free path of a DM particle in the DM gas. Using the definition of $\lambda_\chi$ as 
\begin{equation}
    \lambda_\chi = \frac{1}{n_\chi \sigma_{\chi \chi}}~,
\end{equation}
where $n_\chi$ is the DM number density and $\sigma_{\chi \chi}$ is the dark matter self scattering cross section, we can make a statement about the the maximum self scattering cross section in this limit: 
\begin{equation}
    \sigma_{\chi\chi} \leq \frac{1}{10 \;n_\chi \;R_\star} \simeq 10^{-22} \text{ cm}^2 \left(\frac{m_\chi}{1 \text{ GeV}}\right)\left(\frac{10^{10}\text{ GeV/cm}^{3}}{\rho_\chi}\right)\left(\frac{R_\odot}{R_\star}\right)~,
\end{equation}
where $\rho_\chi$ is the DM mass density. In the case that the probability for dark matter scattering off the object is less than 1, the radius of the object can be replaced with the appropriate length scale $R_\star \sim \sqrt{A_{\rm eff}}$, with $A_{\rm eff}$ the effective area for scattering defined in Eq.~\eqref{eq: A_eff}.

\section{Dark Matter Velocity Near Sagittarius A$^\star$}
\label{app: DM velocity}
Throughout this study we work in the limit that the velocity of the object in question is large, such that the velocity of the surrounding dark matter can be ignored. We will discuss here the applicability and limitations of this assumption. Let us first discuss why it is reasonable to at least assume that the DM velocity never dominates over the velocity of the objects we consider. The basic argument comes from the fact that we always consider the effects of a dark drag force only during the pericenter passage of an object's orbit. As the majority of the surrounding dark matter at that point is not likely to also be at a pericenter phase of its orbit, the average velocity is expected to be lower. This can be seen explicitly from the fact that the DM velocity distribution around a super-massive black hole is expected to take the form of a Maxwell-Boltzmann distribution:
\begin{equation}
    \label{eq: DM vel dist}
    f_\chi(\vec{v}) = \frac{1}{v_\chi^{3}\pi^{3/2}}e^{-\vec{v}^2/v_\chi^2}~,
\end{equation}
with characteristic velocity 
\begin{equation}
\label{eq: DM vel avg}
    v_\chi = \sqrt{\tfrac{G M_{\rm BH}}{r}}~,
\end{equation} 
where this simple estimate is obtained from assuming Keplerian orbits and extrapolating the last measured velocity dispersion towards the Galactic Center \cite{2013PASJ...65..118S}. Comparing Eq.~\eqref{eq: DM vel avg} to the pericenter velocity of an object with eccentricity $e$ at the same distance
\begin{equation}
    v_p = \sqrt{\frac{GM_{\rm BH}(1+e)}{r}}~,
\end{equation}
one sees that the pericenter velocity dominates. The one exception to the rule that we only consider drag at pericenter is the full Velocity Verlet simulation of giant branch stars in Sec. \ref{sec:miss_giants}. The full simulation considers drag induced at all points of the orbit, but as discussed in the relevant section, effectively all of the energy dissipation happens during pericenter anyway, and the apocenter plays a very subdominant role.

With the understanding that the DM velocity is effectively always subdominant to the orbital velocity in question, let us now estimate how good of an approximation it is to set $v_\chi = 0$. As mentioned in Sec. \ref{subsec: drag total reflection}, taking into account the ambient dark matter velocity leads to the drag force in the geometric limit given by Eq. \eqref{eq: Fdrag full}, which assumes specular scattering, where dark matter scatters off of the body in question at an angle equal to the angle of incidence. However, Ref.~\cite{Baines:1965} showed that the exact same formula results from assuming the DM was instantaneously captured, and hence imparts momentum $m_\chi \vec{v}_{\rm rel}$ with each scatter. Interestingly, this is the same average momentum transfer that occurs in the isotropic limit of light dark matter scattering, in the partial reflection regime. As such, the derivation in Ref.~\cite{Baines:1965} can be carried out, with each collision weighted by an average probability for scattering given by 
$\frac{N_{\rm SM}\times \sigma_{\chi_T} }{\pi R_\star^2}$, to yield Eq. \eqref{eq: Fdrag full} with $\pi R_\star^2$ replaced with $A_{\rm eff} = N_{\rm SM}\times \sigma_{\chi T}$. In the heavy limit, the same can be said if we replace $m_\chi$ with $m_T$ in the momentum transfer. Effectively, this means that taking into account the non-zero, dark matter velocity dispersion leads to a drag force in the partial reflection regime given by 

\begin{equation}
    \label{eq: Fdrag full isotropic}
    \vec{F}_{\rm drag} = \begin{cases} -\left(\frac{m_{T}}{m_\chi}\right)A_{\rm eff} \;2n_\chi \left[T_\chi \, \frac{8v^2}{3\sqrt{\pi}v_\chi^2}\left(\left(\frac{v_\chi}{v}\right)^2+\frac{9\pi}{64}\right)^{1/2}\right] \, \hat{v}~ \, \quad m_\chi \gtrsim m_{T} \\\\
    -A_{\rm eff} \;2n_\chi \left[T_\chi \, \frac{8v^2}{3\sqrt{\pi}v_\chi^2}\left(\left(\frac{v_\chi}{v}\right)^2+\frac{9\pi}{64}\right)^{1/2}\right] \, \hat{v}~  \qquad \qquad  m_\chi \lesssim m_{T}
    \end{cases},
\end{equation}
which of course reduces to Eq.~\eqref{eq: Fdrag eff} in the limit that $v \gg v_\chi$. As mentioned in Sec. \ref{subsec: drag total reflection}, it is therefore fully conservative to assume the that dark matter is at rest in both the geometric and partial reflection regimes. Still, to get a sense for how large deviations from the true force are when taking $v_\chi =0$, consider the case that $v_\chi = 0.5 v$. Plugging in this value to Eq.~\eqref{eq: Fdrag full isotropic}, one finds a force that is only $\sim25\%$ larger than the force given by Eq.~\eqref{eq: Fdrag eff}. It is therefore not only conservative, but also entirely reasonable to set $v_\chi=0$ given the uncertainty inherent to all of our results.

The final situation that we have not discussed regards the non-isotropic scattering with heavy nuclei in Sec. \ref{sec:heavy_inel_DM}. In this case, the average momentum transfer in a given collision is not simply $m_T \vec{v}_{\rm rel}$, but rather a more complicated function dependent on the Helm-averaged energy transfer in Eq.~\eqref{eq:average_recoil}. The naive expectation would be that the drag force in this limit deviates to a similar extent as in the isotropic case, when setting $v_\chi = 0$. However, the preference for forward scattering could change the picture more qualitatively. We leave an investigation of non-zero dark matter velocity effects on this regime to future work, but note that for Helium, scattering is nearly isotropic.

\section{Momentum Transfer Angle}
\label{App: f theta_c}
We review here the approximations we have made to the drag force in the inelastic regime analyzed in the main text. In general terms, the drag force can be expressed as an integral over the momentum transfer rate,
\begin{equation}
    | \vec{F}_{\rm drag} | = \frac{\rho_\chi}{m_\chi}N_{\rm SM} \, \bigg| \int \vec{v}_\chi \, f(\vec{v}_\chi) \,d^3\vec{v}_\chi \int \frac{d\sigma}{dE_R} (\vec{q} \cdot \hat{v}_\chi) \, dE_R \bigg|~,
    \label{eq:drag_force_appendix}
\end{equation}
where $N_{\rm SM}$ is the number of SM targets in G2, $f(\vec{v}_\chi)$ is the DM's velocity distribution in the rest frame of G2, $\vec{q}\cdot \hat{v}_\chi$ represents the meaningful piece of the momentum transfer to the drag force (momentum transfer in all lateral directions averages to zero over many scatters), and the differential cross-section is 
\begin{equation}
    \frac{d\sigma_{\chi N}}{dE_R} = \frac{\sigma_{\chi n} \, m_N}{2 v_\chi^2 \mu_{\chi n}^2} |F_{\rm helm}(E_R)|^2.
\end{equation}
Eq.~\eqref{eq:drag_force_appendix} can be further simplified as we are neglecting the DM's velocity dispersion in our calculations, in which case we can approximate the distribution as $f(\vec{v}_\chi) \propto \delta(\vec{v_\chi} - \vec{v})$ where $\vec{v}$ is G2's pericenter velocity. In other words, in G2's rest frame, all incoming DM particles share the same velocity given by how fast the cloud was moving during pericenter passage. Thus, we obtain,
\begin{equation}
    \label{eq: app C drag parametric}
    | \vec{F}_{\rm drag} | = \frac{\rho_\chi}{m_\chi}N_{\rm SM}\;v \times \int \frac{d\sigma}{dE_R} (\vec{q} \cdot \hat{v}) \, dE_R,
\end{equation}
where $\rho_\chi$ and $v$ are understood to be evaluated at pericenter. The projected momentum transfer can be calculated directly from energy conservation,
\begin{equation}
    E_f = m_N + (m_\chi+\delta m_\chi) + \frac{(\vec{p}-\vec{q})^2}{2(m_\chi + \delta m_\chi)} + \frac{\vec{q}^2}{2 m_N} = E_i = m_N + m_\chi+ \frac{\vec{p}^2}{2 m_\chi}~,
\end{equation}
which yields the following equality when neglecting terms of order $\delta m_\chi \ll m_\chi$
\begin{equation}
    \frac{\vec{p} \cdot \vec{q}}{m_\chi} = \delta m_\chi + \frac{\vec{q}^2}{2 \mu_{\chi N}}~.
\end{equation}
This means that the projected momentum transfer is
\begin{equation}
    \vec{q} \cdot \hat{v} = \frac{1}{m_\chi v} (\vec{q} \cdot\vec{p}) = \frac{1}{v} \left(\delta m_\chi + \frac{\vec{q}^2}{2 \mu_{\chi N}}\right) = \frac{1}{v}\left[\delta m_\chi + \left(1 + \frac{m_N}{m_\chi}\right)E_R\right],
\end{equation}
where by definition the recoil energy is $E_R = \mathbf{q}^2/2m_N$. Plugging this form of the projected momentum transfer into the integral in Eq. \eqref{eq: app C drag parametric} (and ignoring constant factors) yields
\begin{equation}
\begin{split}
    | \vec{F}_{\rm drag} | &\propto \int_{E_{\rm min}}^{E_{\rm max}} |F(E_R)|^2 \; (\vec{q}\cdot \hat{v}) \, dE_R = \int_{E_{\rm min}}^{E_{\rm max}} |F(E_R)|^2 \frac{1}{v}\left[\delta m_\chi + \left(1 + \frac{m_N}{m_\chi}\right)E_R\right] \, dE_R\\\\
    &= \frac{1}{v}\left[\delta m_\chi + \left(1 + \frac{m_N}{m_\chi}\right)\langle E_R\rangle \right]\int_{E_{\rm min}}^{E_{\rm max}} |F(E_R)|^2 \, dE_R = (\vec{q}\cdot\hat{v})\bigg|_{E_R = \langle{E_R}\rangle} \times\int_{E_{\rm min}}^{E_{\rm max}} |F(E_R)|^2 \, dE_R,
\end{split}
\end{equation}
using our average recoil energy definition, $cf.$ Eq.~\eqref{eq:average_recoil}. Furthermore, it is straightforward to verify that
\begin{equation}
    \int_{E_{\rm min}}^{E_{\rm max}} |F(E_R)|^2 \, dE_R = |F(\langle E_R \rangle)|^2 + \mathcal{O}\left(\left(\frac{E_{\rm max} + E_{\rm min}}{E_N}\right)^2\right).
\end{equation} 
It is therefore valid to leading order in $(E_{\rm max} + E_{\rm min})/E_N$ to write the total drag force in Eq. \eqref{eq: app C drag parametric} as
\begin{equation}
\label{eq: drag nuclear leading order}
   | \vec{F}_{\rm drag} | \approx \frac{\rho_\chi}{m_\chi} \, N_{\rm SM} \; v\;\overline{\sigma}_{\chi T}\, (\overline{\vec{q}\cdot \hat{v}}) \,,
\end{equation}
where $\overline{\sigma}_{\chi T}$ is given by Eq.~\eqref{eq:cross_section_nucleus} evaluated at $\langle E_R \rangle$ and $(\overline{\vec{q}\cdot \hat{v}})$ is   $(\vec{q}\cdot \hat {v})$ evaluated at $\langle E_R \rangle$. With this in mind, we can re-write Eq.~\eqref{eq: drag nuclear leading order} in terms of the center mass scattering angle associated with the average recoil energy $\theta^*_c$, in order to write $F_{\rm drag}$ in a manner that smoothly connects to the isotropic case.

Since the maximum velocities we consider for objects around Sgr.~A$^\star$ are of order $\sim 10^{-2}$, we will work in the non-relativistic limit. Moreover, we will work in the limit where DM is much heavier than the target. The momentum of either particle in the center of mass frame prior to the collision is simply
\begin{equation}
    \vec{p}_{\rm CM} = \pm \, \mu_{\chi T} \, \vec{v},
\end{equation}
where $\mu_{\chi T}$ is the reduced mass of the DM particle and the target, and $\vec{v}$ is given by G2's pericenter velocity. For elastic collisions, $\vec{p}_{\rm CM}$ of course maintains magnitude and is rotated by some angle. However, if the collision is inelastic, the center of mass momentum both rotates and changes magnitude post-collision. Denoting this new vector by $\vec{p}_{\rm CM}^\prime$, its magnitude is \cite{Alvarez:2023fjj}
\begin{equation}
    |\vec{p}_{\rm CM}^{\, \prime}|^2 = |\vec{p}_{\rm CM}|^2 - \frac{\left((m_\chi + \delta m_\chi)^2 - m^2_\chi\right) \left(2 E^2_{\rm CM} + 2 m_T^2 - m_\chi^2 - (m_\chi + \delta m_\chi)^2\right)}{4 E^2_{\rm CM}}~,
\end{equation}
where $E_{\rm CM}$ is the center of mass energy, and $\delta{m_\chi}$ is the mass splitting of the DM state. Note that the above is a fully relativistic expression, although for our case the second term on the right hand side can be simplified as we are always in the limit $\delta m_\chi \ll m_\chi$. We write the full vector as $\vec{p}_{\rm CM}^{\, \prime} = |\vec{p}_{\rm CM}^{\, \prime}| \, \hat{n}_0$, where $\hat{n}_0$ is the unit vector that determines the direction relative to the initial $\vec{p}_{\rm CM}$, and is therefore related to the center of mass angle $\theta_c$. 

In the non-relativistic limit, $\vec{q} = \vec{q}_{\rm CM} = \vec{p}^{\, \prime}_{\rm CM} - \vec{p}_{\rm CM}$ is the transferred momentum. We are interested in the projection $\vec{q} \cdot \hat{v}$, where $\hat{v}$ is the unit vector denoting the relative velocity, as momentum transfer in any other direction averages out to zero over multiple scattering events. In other words, we want the parallel component of the momentum transfer, as this is what contributes to the drag.

Using the unit vector definition $\hat{n}_0$ we get a new expression for $\vec{q}\cdot \hat{v}$ in terms of $\theta_c$:
\begin{equation}
    \vec{q} \cdot \hat{v} = |\vec{p}^{\, \prime}_{\rm CM}| \cos(\theta_c)- |\vec{p}_{\rm CM}|~.
\end{equation}
We define then the function $f(\theta^*_c)$  as the ratio
\begin{equation}
    f(\theta^*_c) = \frac{|\vec{p}^{\, \prime}_{\rm CM}| \cos(\theta^*_c)- |\vec{p}_{\rm CM}|}{|\vec{p}^{\, \prime}_{\rm CM}| \cos(\pi/2)- |\vec{p}_{\rm CM}|} = 1 - \frac{|\vec{p}^{\, \prime}_{\rm CM}|}{|\vec{p}_{\rm CM}|}\cos(\theta^*_c)~,
\end{equation}
normalized to 1 for scattering angle $\theta_c =\pi/2$. Here $\cos(\theta^*_c)$ is given by the helm-averaged value $\langle \cos(\theta_c) \rangle$, computed from the average energy transfer, Eq.~\eqref{eq:average_recoil}. $f(\theta^*_c)$ evaluates to unity in the limit of isotropic, elastic scattering, but for heavy inelastic DM, where form factor effects limit the scattering angle, $f(\theta^*_c) < 1$ generically. Finally, note that for a maximally inelastic collision with $\delta m_\chi = \delta m_\chi^{\rm max}$, $f(\theta^*_c) \rightarrow 1$ since $|\vec{p}^{\, \prime}_{\rm CM}|$ vanishes while $\cos(\theta^*_c)$ saturates to 1. 

With this definition of $f(\theta^*_c)$ we can write the total drag force as

\begin{equation}
    |\vec{F}_{\rm drag}| =  f(\theta^*_c) \, \left(\frac{m_{T}}{m_\chi}\right)\rho_\chi \, A_{\rm eff}(\theta^*_c) \, v^2~,
\end{equation}

where $A_{\rm eff}(\theta^*_c) = N_{\rm SM}\times \overline{\sigma}_{\chi T}$, which takes a similar form to the isotropic case.

\section{The Shape of G2 as a Gas Cloud}
\label{app: G2 shape}

Although we model G2 as a sphere, in practice, during pericenter passage, G2 can be tidally stretched along the orbital plane and compressed along the axis perpendicular to the orbital plane \cite{Morsony:2015uqa,Steinberg2018}. However, as long as the mean free path of DM remains longer than the size of the object (i.e. we remain in the partial reflection regime), the observed acceleration can be mapped to the same dark matter parameter space because $A_{\rm eff}$ is independent of the object's shape in this regime.
Given that the exact dimensions of G2 during pericenter are not well known, here we estimate how much G2 can vary from a spherical shape before we are no longer in the optically thin regime. The condition for which G2 remains optically thin to DM scattering is given by
\begin{equation}
\label{eq:mean_free_path}
    \lambda = \frac{1}{n_{\rm SM} \sigma_{\chi T}} \gtrsim L,
\end{equation}
where $\lambda$ is the mean free path of a DM particle in G2 and $n_{\rm SM}$ is the number density of SM particles in G2. Here, we model G2 as roughly cylindrical at pericenter with length $L$ and circular end caps of area $A$. Assuming that $n_{\rm SM}$ is uniform and that the total number of particles in G2 remains constant throughout its tidal evolution, we can use our earlier definition of $N_{\rm SM} = M_{\rm G2}/m_n$ to obtain the number density $n_{\rm SM} = N_{\rm{SM}}/LA$. This leads to an approximate lower bound on $A$,
\begin{equation}
\label{eq: G2 A min}
    A \gtrsim 10^{28} \ \rm{cm}^2 \left(\frac{\sigma_{\chi T}}{10^{-24} \rm{cm}^2}\right) \left(\frac{N_{\rm{SM}}}{10^{52}}\right),
\end{equation}
for which we remain in the optically thin regime. For comparison, the cross-sectional area of G2 assuming a spherical shape is $A_{\rm{sphere}} = \pi R_{\rm{G2}}^2 \sim 10^{31} \ \rm{cm}^2$. This implies that for most cross sections and DM profiles considered in this work, G2's cross sectional area can decrease by multiple orders of magnitude without violating our assumption of being in the optically thin regime. 

In principle, it is possible for G2 to compress even further beyond the point where our assumptions hold ~\cite{Steinberg2018}. However, such a scenario is entirely model dependent as the existence of a magnetic field flowing through G2 can generate a magnetic force that helps resist collapse from tidal forces and pressure from ambient gases~\cite{Shcherbakov_2014}. In this model, the cross-sectional area of G2 during pericenter can stabilize at $A \sim 10^{29} \ \rm{cm}^2$. Even in the absence of a magnetic field, Ref.~\cite{Shcherbakov_2014} predicts that the cross-sectional area of G2 decreases by a factor of only about 2500 at pericenter, rather than the extreme compression predicted by \cite{Steinberg2018}. This discrepancy highlights the sensitivity of G2's pericenter shape to cooling assumptions.

Finally, we note that the lower bound on $A$ in Eq.~\eqref{eq: G2 A min} relies on G2 retaining a uniform density and relatively simple shape, which in general may not be true. Eq.~\eqref{eq: G2 A min} should therefore be taken as evidence for why our assumptions are reasonable, rather than a robust proof, which requires a more detailed modeling of G2's pericenter evolution that we leave to future work.

\section{Ultralight DM}
\label{sec: ultralight dm}
As one pushes into the ultralight dark matter regime, coherent enhancement of the scattering cross section can become important. In this case, $A_{\rm eff}$ defined in Eq.~\eqref{eq: A_eff} must be corrected, as interactions with a scattering length $\lambda = 1/q$ (for momentum transfer $q$) larger than the interatomic spacing of the object it scatters off will become coherently enhanced \cite{Luo:2024ocg, Gan:2025nlu, Matsumoto:2025rcz}. Taking this effect into account, we can write $A_{\rm eff}$ as
\begin{equation}
\label{eq: Aeff coherent}
    A_{\rm eff}^{\rm coherent} = \sum_{i,j}^{N_{\rm SM}} e^{i \vec{q}\cdot \Delta\vec{r}_{i,j} }\;\sigma_{\chi T}~,
\end{equation}
where $\Delta \vec{r}_{i,j}$ is the displacement between two nucleons or electrons, and $\vec{q}$ is the momentum transfer. 
With this definition, the coherent enhancement can be parameterized by a form factor $F(q)$ as
\begin{equation}
\label{eq:enh}
    \sum_{i,j}^{N_{\rm SM}} e^{i \vec{q}\cdot \Delta\vec{r}_{i,j} } \equiv N_{\text{SM}} + (N_{\text{SM}}^2-N_{\text{SM}})|F(q)|^2. 
\end{equation}
Such a form factor describes the extent of the coherent enhancement at a given momentum transfer, ranging from 0 to 1. The specific functional form of $F(q)$ will depend on the assumed shape and density distribution of G2. The sensitivity to ultralight dark matter will therefore be more dependent on the shape even in the optically thin regime, unlike the other dark matter masses that we explore (see App.~\ref{app: G2 shape}). Given the uncertain nature of G2's shape around pericenter, we leave an exploration of the ultralight DM regime to future work.


Furthermore, for light enough masses, there are other effects that may become relevant regarding the orbital kinematics of G2, compared to the simple momentum transfer used in this work. For example, there are strong limits that can be set on ultralight dark matter candidates which couple to the Standard Model by considering additional acceleration induced by classical field gradients or quantum force law corrections \cite{Banerjee:2022sqg}. It would be interesting to see if analogous classical field effects on G2 could have a strong impact on its orbital evolution. One could also consider the possibility of density enhancements due to boson clouds radiated from Sgr.~A$^\star$ (see e.g.~\cite{Dolan:2007mj, Arvanitaki:2010sy,Brito:2015oca,Bai:2025yxm}), which could be examined in the context of G2's orbital decay.

\bibliographystyle{apsrev4-1} 
\bibliography{refs}

\end{document}